\documentclass[english,aps,prper,reprint,showpacs,titlepage,longbibliography]{revtex4-2}   

\usepackage{soul}
\usepackage{xcolor}
\soulregister{\textit}{1}

\definecolor{lightgray}{gray}{0.9}
\usepackage{listings}
\sethlcolor{lightgray}
\usepackage[utf8]{inputenc}
\usepackage[compatibility=false]{caption} 
\usepackage{rotating} 
\usepackage{booktabs} 
\usepackage{tabularx} 
\usepackage{ragged2e} 
\usepackage{caption} 
\usepackage[T1]{fontenc}	
\usepackage{geometry}
\usepackage{times}
\usepackage{hyperref}  
\hypersetup{colorlinks=true,urlcolor=blue,citecolor=blue,linkcolor=blue} 
\usepackage{array}
\usepackage{enumerate}
\usepackage{enumitem}
\usepackage{amsmath}
\usepackage{caption}
\usepackage{amssymb}
\usepackage{tikz}
\usepackage{graphicx}
\usepackage{rotating}
\usepackage{caption}
\usepackage{placeins}
\usepackage{multirow}
\usepackage{adjustbox} 
\usepackage{tcolorbox}
\usepackage{ragged2e}
\usepackage{float}
\usepackage[utf8]{inputenc}

\usepackage[normalem]{ulem}
\usepackage{setspace} 
\usepackage{placeins} 
\usepackage{adjustbox} 
\begin{document}
\begin{titlepage}

\title{Applying a STEM Ways of Thinking Framework for Student-generated Engineering Design-based Physics Problems}

 \author{Ravishankar Chatta Subramaniam}
 \affiliation{Department of Physics and Astronomy, Purdue University, West Lafayette, IN 47907, U.S.A.} 

 \author{Nikhil Borse}
 \affiliation{Department of Physics and Astronomy, Purdue University, West Lafayette, IN 47907, U.S.A.} 
 \author{Winter Allen}
 \affiliation{Department of Physics and Astronomy, Purdue University, West Lafayette, IN 47907, U.S.A.} 
 
 \author{Amogh Sirnoorkar}
 \affiliation{Center for Advancing the Teaching and Learning of STEM, Purdue University, West Lafayette, IN 47907, U.S.A. } 
 \author{Jason W. Morphew}
 \affiliation{School of Engineering Education, Purdue University, 610 Purdue Mall, West Lafayette, IN 47907, U.S.A.}
   
 \author{Carina M. Rebello}
 \affiliation{Department of Physics, Toronto Metropolitan University, 350 Victoria Street, Toronto, ON M5B 2K3, Canada}
  
 \author{N. Sanjay Rebello}
 \affiliation{Department of Physics and Astronomy, and  Department of Curriculum and Instruction, Purdue University, West Lafayette, IN 47907, U.S.A.} 

\keywords{}

\begin{abstract}

This second paper in a multi-part series builds on the first, which introduced the Ways of Thinking for Engineering Design-based Physics (WoT4EDP) framework for STEM education. In this paper, we apply the framework to analyze student-groups' discussion transcripts and written reports as they generated and solved an engineering design (ED) problem in an introductory physics laboratory. We present our findings through a detailed qualitative analysis, drawing extensively from the qualitative methods literature. We outline our analytical approach, present coding charts, and employ qualitative methods such as thematic analysis and thick description to convey our findings. Furthermore, in line with the recent recommendation in qualitative research literature, we include a reflexivity statement. In doing so, we contribute to ongoing efforts to enhance the rigor in qualitative physics education research. We examine: (i) the design and science aspects students address in their problem statements; (ii) how they engage in design-, science-, and mathematics-based thinking, as well as their metacognitive reflection while developing solutions; and (iii) how they incorporate computational thinking. Our results  highlight the need for: (i) increased guidance for iterative problem framing; (ii) structured support for assessing design limitations, engaging in a feasibility study, adopting a systematic approach to applying physics and mathematics in their iterations, and engaging in  metacognitive reflection; and (iii) integration of computation-based activities into laboratory tasks with appropriate scaffolding. Based on our analysis, we provide valuable insights for educators and researchers in designing physics-based engineering design tasks and promoting interdisciplinary problem-solving in STEM education.

    \clearpage
  \end{abstract}

\maketitle
\end{titlepage}
\maketitle
\section{Introduction}

Understanding students' thinking in authentic science, technology, engineering, and mathematics (STEM) learning environments by integrating multiple disciplines is essential for enhancing educational practices and outcomes~\cite{kelley2016conceptual, roehrig2021beyond, geesa2020integrative}. Investigating how students approach~\textit{engineering design} (ED)-based tasks~\cite{ravi_aapt_2024, ravi_perc_2024, ravi_SWoT_01_2024, ravi_perc_2023, capobianco2013shedding} provides valuable insights into their problem-solving strategies and the integration of various disciplines~\cite{ravi_prper_2024}. Recently, the concept of~\textit{Ways of Thinking} (WoT)~\cite{ravi_SWoT_01_2024} has emerged as an effective framework for examining interdisciplinary problem-solving in educational settings. 

\begin{figure}[!htbp]
\caption{The~\textit{WoT4EDP} Framework, reproduced from the first paper~\cite{ravi_SWoT_01_2024} of this two-part series. The double-headed arrows illustrate the interconnectedness of each element with the other four~\textit{Ways of Thinking}. While this study does not explicitly focus on the interconnectedness of the elements, there will be instances where the connections becomes evident. }
\fbox{\includegraphics[width=\linewidth]{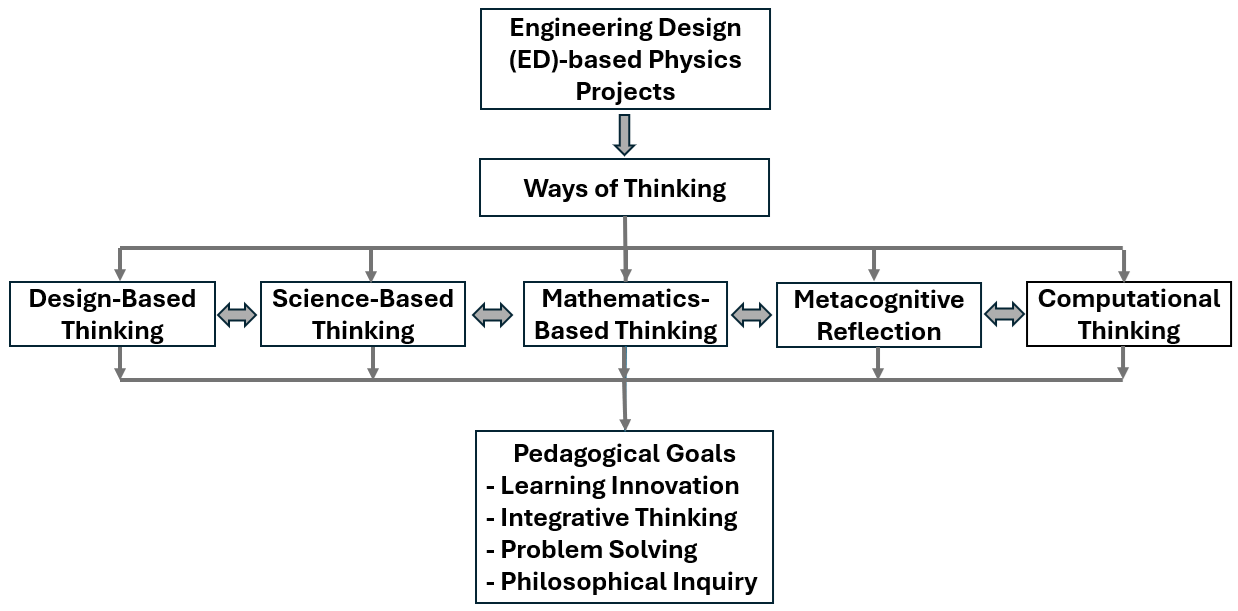}}
\label{fig:wot_framework}
\end{figure}

In the first paper~\cite{ravi_SWoT_01_2024} of this multi-part series, we introduced the~\textit{Ways of Thinking in Engineering Design-based Physics} (\textit{WoT4EDP}) framework (see Figure~\ref{fig:wot_framework}, and refer Section~\ref{sec:II} for details)  
which integrates five key elements—design, science, mathematics, metacognitive reflection, and computational thinking—tailored specifically for introductory physics laboratory courses based on engineering design. While this paper builds on the first, we provide sufficient background and context to ensure it stands largely on its own. 

While the first paper~\cite{ravi_SWoT_01_2024} was theoretical, this paper is empirical. Our goal here is to demonstrate the application of the~\textit{WoT4EDP} framework to data from student groups’ discussions and written reports. These~\textit{artifacts}~\cite[p.17]{otero2009getting}) stem from a student-generated engineering design (ED) problem~\cite[p.140]{ravi_aapt_2024} within the laboratory component of a calculus-based introductory, first semester physics course. While the first paper~\cite{ravi_SWoT_01_2024} introduced the~\textit{WoT4EDP} framework and highlighted its significance in undergraduate STEM education, this paper describes a qualitative approach to apply the~\textit{WoT4EDP} framework to uncover how students engage with its five elements while tackling student-generated engineering design challenges. By utilizing this framework, we aim to discern patterns in students' thinking and problem-solving strategies, contributing to improved practices in STEM education. Although the five elements are interconnected, we discuss them separately in this study. While we occasionally point out instances of interconnectedness as a prelude to our next paper, a more detailed analysis will be presented in the subsequent installment of this multi-part series.

We investigate the following  research questions focusing on student-generated engineering design (ED) problems. 

\begin{enumerate}
\item[] {\bf RQ1:} {\em What do student teams’ engineering design problem statements reveal about the design and science aspects they consider?}  \label{RQ1}

\item[] {\bf RQ2:} {\em How do student teams engage with the elements of design, science, mathematics, and metacognitive reflection within the context of a student-generated engineering design problem?}  \label{RQ2}

\item[] {\bf RQ3:} {\em In what ways do students demonstrate computational thinking specifically related to Python coding as they tackle engineering design challenges?}  \label{RQ3}
\end{enumerate}

The first question~\hyperref[RQ1]{RQ1} examines how students articulate their problem statements. The second question~\hyperref[RQ2]{RQ2} explores how design, science, mathematics, and metacognitive thinking manifest in student teams' work. Given the emphasis on computational tasks in physics courses and the fact that computational thinking (CT) is a novel element in our~\textit{WoT4EDP} framework—absent in contemporary frameworks on STEM ways of thinking~\cite{dalal2021developing, english2023ways, slavit2022analytic}—we dedicate the third question~\hyperref[RQ3]{RQ3} to exploring CT. 

While applying the~\textit{WoT4EDP} framework is the primary goal of this empirical study, we also take this opportunity to discuss key methodological challenges in qualitative research. Through this study, we demonstrate the value of qualitative methods in capturing the complexities of student thinking and problem-solving within physics education research (PER)
~\cite{roberts1982place, devetak2010role, stanley2024qualitative}. This work, like our earlier studies~\cite{ravi_prper_2024, ravi_SWoT_01_2024}, is deeply rooted in and extends the rich tradition of qualitative research within PER~\cite{otero2009getting, hammer1995student, kuo2013students, bakhtinian_2023, phy_identity_gonsalves_2023, research_design_cohen_2023, sirnoorkar2024student,sirnoorkar2023sensemaking, sirnoorkar2021methodology,sirnoorkar2023analyzing, video_euler_2023, phen_graphy_guisanola_2023, Ai_qlr_tschisgale_2023, stud_resources_2023, mcpadden_2023, epistemic_beliefs_watson, redish2015language, rachel_scherr_2008}. In doing so, we extend the discussion on the development and application of WoT frameworks~\cite{english2023ways, slavit2021student, dalal2021developing, ravi_SWoT_01_2024}. 

This manuscript is structured as follows: In Section~\ref{sec:II}, we briefly outline the elements of the~\textit{WoT4EDP} framework. Section~\ref{sec:III} reviews the background literature on authentic learning experiences, their connections to~\textit{WoT4EDP}, and qualitative research in PER, not only situating our study but also grounding its methodological approach. We then present the study’s context, along with methods of data collection and analysis, in Section~\ref{sec:IV}. Section~\ref{sec:V} presents the findings and discussion, where we address our research questions. In Section~\ref{sec:VI}, we identify the study's limitations. We summarize our findings in Section~\ref{sec:VIII}, outline implications in Section~\ref{sec:IX}, and discuss future work in Section~\ref{sec:X}. While we discuss methodology throughout the paper, Section~\ref{sec:VII} is dedicated to a reflexivity statement in line with recent trends. The coding sheets in the appendixes reinforce our commitment to transparency.  

\section{The W\texorpdfstring{\lowercase{o}}{o}T4EDP Framework}\label{sec:II}

The \textit{Ways of Thinking for Engineering Design Problems}~\textit{WoT4EDP} framework (Figure~\ref{fig:wot_framework}), introduced in the first paper~\cite{ravi_SWoT_01_2024} of this multi-part series, provides a structured perspective for analyzing the diversity of student thinking approaches when addressing engineering design-based physics problems. To ensure continuity and provide necessary background, we briefly summarize the framework here before applying it in this study. 
The framework consists of the following five interconnected elements:

\begin{enumerate}
    \setlength{\itemsep}{1.5pt} 
    \setlength{\parskip}{0pt}  
    \item \textit{Design-Based Thinking} (DBT) (also  referred as design thinking).
    
    \item \textit{Science-Based Thinking} (SBT) (also  referred as science thinking or scientific thinking).
    
    \item \textit{Mathematics-Based Thinking} (MBT) (also  referred as  mathematical thinking).
    
    \item \textit{Metacognitive Reflection} (MER), and
    
    \item \textit{Computational Thinking} (CT).
\end{enumerate}

These dimensions collectively offer a comprehensive lens for examining how students integrate creativity, scientific principles, mathematical reasoning, reflective practices, and computational skills in their approaches to problem-solving. We briefly describe these elements and explain why each element is relevant to this study in Table~\ref{tab:wot_description}. 

\renewcommand{\arraystretch}{1.2} 
\begin{table*}[!htbp]
\begin{center}
\captionsetup{justification=raggedright} 
\caption{Elements of the~\textit{WoT4EDP} framework as applied to our context in this study. 
Educators and researchers must feel empowered to modify and apply these descriptors as may fit their local contexts and pedagogical goals. More details follow in the later sections.
DBT--Design Based Thinking; SBT--Science Based Thinking; MBT--Mathematics Based thinking; MER--Metacognitive Reflection; CT - Computational Thinking.}
\label{tab:wot_description}
\begin{ruledtabular}
\begin{tabular}{p{0.08\linewidth} p{0.5\linewidth} p{0.4\linewidth}} 
    \textbf{SWoT Element} & \textbf{Description} & \textbf{Why It Matters} \\
    \hline
    DBT & Frame the problem; identify criteria and constraints; identify stakeholders; brainstorm multiple solutions; iterate, select the best solution; consider design aspects (dimensions, mechanism); provide metrics; outline trade-offs; develop solution (product / process); identify design limitations; conduct feasibility study; communicate. & Encourages creative problem-solving, iterative thinking, and decision-making under constraints, preparing students for real-world design challenges, engage in collaborative work; adopt multiple perspectives. \\ 
    SBT & Apply fundamental physics principles (Momentum, Energy, and Angular Momentum) and related concepts in analyzing and refining designs; integrate physics into solutions; outline assumptions and approximations; conceptual correctness; qualitative reasoning; paying attention to significant digits and units while presenting numerical results. & Strengthens scientific reasoning, ability to analyze data, reinforces core physics concepts, and fosters the ability to justify design choices with physics-based arguments. \\     
    MBT & Use mathematical concepts (for eg., algebra, geometry, trigonometry, and calculus) to analyze design problems; perform calculations; derive / establish relationships; consider numerical aspects; identify variables and constants; utilize graphs and tables to detect or depict patterns and trends; engage in math-based reasoning. & Develops analytical and quantitative skills, allowing students to model and optimize solutions mathematically; enhances their ability to solve complex problems, fostering skills in recognizing relationships and abstract concepts vital for STEM problem-solving. \\ 
    MER & Reflect on design decisions, iterations, and the integration of science, math, and Python code to improve approaches; engage in reflective evaluation of their solution. & Enhances students’ ability to evaluate their own learning, adapt strategies, and make informed improvements in design and reasoning. \\ 
    CT & In this paper we restrict to the application computational techniques using Python programming (CT-lite, one could say), to simulate, model, and analyze design challenges; ability to break a complex problem into simpler parts; use coding to integrate physics and math into problem-solving. & Equips students with computational tools to handle complex problems, automate analysis, and simulate real-world systems, making solutions more robust. \\
        
\end{tabular}
\end{ruledtabular}
\end{center}
\vspace{-0.5cm} 
\end{table*}

It is important to recognize that these five elements do not always function in isolation. Students’ thinking is inherently fluid and multidimensional, making it difficult to categorize neatly into discrete silos~\cite{ravi_prper_2024}. The overlap among these elements will be evident as students frequently draw on multiple elements simultaneously. For example, they may use Python to model a physics concept, incorporating mathematical tools, while refining their design iteratively based on scientific reasoning. This interconnectedness warrants a deeper analysis, which is why we expanded this study into a multi-part series rather than the originally planned two-part series~\cite{ravi_SWoT_01_2024}.

Despite this overlap, educators can still benefit from focusing on each element individually to help students develop a deeper awareness of how different ways of thinking contribute to problem-solving. This intentional emphasis strengthens students’ ability to navigate and integrate various approaches, ultimately equipping them with the skills necessary to tackle complex, real-world problems.

As to the pedagogical goals our framework may support, the list we have presented is neither rigid nor exhaustive. Our view is that these are broad goals, and educators must feel empowered to tweak the goals to fit their local contexts and interests~\cite{ravi_SWoT_01_2024}. 

Before proceeding, we note that there is considerable debate on the definition of~\textit{engineering design}. The literature is replete with terms such as `ill-structured', `ill-defined', or even `wicked' problems~\cite{ravi_SWoT_01_2024, ravi_prper_2024}. For further discussion, we refer the reader to an insightful article by Grubbs and Strimel~\cite{grubbs2015engineering}. In this study, we define an engineering design problem as a real-world challenge engineers might face, though professional tasks are typically more complex. The problems discussed here serve as simulations for first-year engineering students. Our students selected problems where physics principles—momentum, energy, and angular momentum—apply, aligning with the flexible~\textit{WoT4EDP} framework. Since students framed their own problems, rigid expectations could frustrate them and detract from the task's overall objectives. Thus, we take a pragmatic approach: any student-identified engineering design problem that integrates physics, mathematics, and Python coding is considered sufficient. 

\section{Literature Review}\label{sec:III}

The first paper~\cite{ravi_SWoT_01_2024} of this series focused on the theoretical foundations of our~\textit{WoT4EDP} framework, with the literature review tailored accordingly. In this second, empirical paper, we situate our work within broader literature encompassing authentic learning experiences, WoT frameworks, physics education research, and qualitative research methodologies. In this section, we first examine the role of authentic learning experiences, within which we position our approach of ED-based physics laboratory tasks (Section~\ref{sec:III.A}). Next, we explore the potential of frameworks such as~\textit{WoT4EDP}, introduced in the first paper~\cite{ravi_SWoT_01_2024}, in supporting student thinking in ED-based tasks (Section~\ref{sec:III.B}). Finally, we position our work within ongoing efforts to advance the application of qualitative research methodologies in PER (Section~\ref{sec:III.C}).

\subsection{Authentic Learning Experiences}\label{sec:III.A}

The National Research Council's 2012 report on~\textit{Discipline-Based Education Research} makes an important observation that effective instruction involves~\textit{``having students work in groups, and incorporating authentic problems and activities''}~\cite[p.3]{nrc_dber_2012}. In our context of physics education, this is further amplified by Watkins {\em et al.} who emphasize that to further interdisciplinary learning in physics classrooms, we need to explore constructive ways of thinking and organize our understanding of the scientific disciplines. Relevant to this study is their characterization of~\textit{authentic} activities in science as:
\begin{quote}
\textit{...those that use tools—such as concepts, equations, or physical tools—in ways and for purposes that reflect how the disciplines build, organize, and assess knowledge about the world.}~\cite[p.2]{redish_watkins_dber} 
\end{quote}

These perspectives on~\textit{authentic} learning and~\textit{ways of thinking} align closely with our focus on real-world Engineering Design-based physics projects and the application of WoT frameworks.

Research demonstrates that embedding Engineering Design (ED) into physics laboratory tasks enhances students' engagement with key science and engineering practices~\cite{NGSS2013, fischer2014interplay, honey2014, NRC2012, NRC2014}. This approach serves to bridge the gap between~\textit{‘`tactics and strategies'’} (ED) and~\textit{‘`practices and the nature of science’'} (Science)~\cite{NGSS2013, ravi_perc_2024}, providing students with opportunities to think like physicists while working on real-world problems. While the integration of Engineering Design in physics learning may be relatively recent, the idea of project-based learning in physics can be traced back to the early 20th century, when Mann introduced the notion of~\textit{``industrial science''}, or more specifically~\textit{``industrial physics''} in our context. He argued that project-based learning was a means to~\textit{``awaken the scientific spirit''} in students~\cite{kapon2018disciplinary, mann1914industrial}. More than a century later, educators and researchers continue to grapple with fundamental questions about what constitutes an~\textit{authentic science}~\cite{chinn2002epistemologically} learning experience, what students gain from authentic inquiry, and how educators can best support and guide students in research-based projects. It is within this ever evolving landscape that we see a meaningful role for our~\textit{Ways of Thinking for Engineering Design-based Physics} (\textit{WoT4EDP}) framework~\cite{kapon2018disciplinary, ravi_SWoT_01_2024, ravi_prper_2024}.

\subsection{Authentic Learning and \textit{WoT4EDP} Framework }\label{sec:III.B}

Montgomery {\em et al.} emphasize that to become~\textit{``expert problem solvers, students must have opportunities to deliberately practice''}~\cite[p.1]{montgomery2024characterizing} problem-solving skills. Our ED-based physics projects align with this perspective: students have the opportunity to develop teamwork skills through group collaboration and engage with real-world scientific and engineering practices (see Table~\cite[p.2]{montgomery2024characterizing}). Students tackle authentic problems that require applying fundamental physics principles, such as momentum, energy, and angular momentum~\cite{chabay2015matter, ravi_prper_2024}. Students use a variety of tools—from writing equations and performing calculations by hand to automating calculations using Python—actively engaging in problem-solving that mirrors real-world contexts. Crucially, students are empowered to choose their own problems of interest, providing them with a~\textit{sense of authorship} or~\textit{agency}~\cite{cavagnetto2020supporting, agency_holmes_2020, kalender2020sense} that can foster deeper engagement and ownership of their learning. 

Merely providing students with opportunities to work on such projects may not always result in achieving the desired learning outcomes unless they are supported by carefully thought out instructional scaffolds~\cite{dasgupta2019improvable, kolod_punt_2003}. Educators must hold high expectations for students while recognizing that deep understanding unfolds step by step. Bliss {\em et al.} capture the need for this balance succinctly:
\begin{quote}
\textit{Teachers need to believe that children can learn
difficult and complex ideas; this is what school is about. But they must be content that often pupils can only do this one step or a few steps at a time.}~\cite[p.60]{bliss1996effective}

\end{quote}

This is precisely where~\textit{Ways of Thinking} (WoT) frameworks play a pivotal role, offering a structured lens to support students as they navigate the gradual process of mastering complex concepts. Evidence suggests that an intentional focus on WoT can significantly enhance students’ learning in STEM environments~\cite{slavit2019stem, slavit2021student, slavit2022analytic, dalal2021developing, english2023ways}. In the first paper~\cite{ravi_SWoT_01_2024} of this multi-part series, we introduced~\textit{WoT4EDP}, a framework designed for ED project-based physics laboratories. This framework draws upon insights from reform documents published by the National Research Council (NRC), the American Association of Physics Teachers (AAPT), the Mathematical Association of America (MAA), the College Board, and other organizations, alongside contemporary research on student learning in STEM~\cite{collegeboard2009science, NGSS2013, AAPT_2023_lab_reco, NRC2011, NRC2012, NRC2014, math_cupm_2015, nrc_meta_2000, nrc_CT_2010}. While our framework is rooted in a specific instructional context, we believe that its principles can be extended to other educational settings.

The role of WoT frameworks in the context of engineering design can be appreciated through the insightful observation of Grubbs and Strimel, who regard engineering design as the~\textit{great integrator}:~\textit{``Because little is known about student cognition during such experiences, future research can provide additional implications and instructional resources to guide implementation''}~\cite[p.87]{grubbs2015engineering}.

\textit{WoT4EDP} focuses on five interconnected~\textit{Ways of Thinking}:~\textit{Design-Based Thinking} (DBT),~\textit{Science-Based Thinking} (SBT),~\textit{Mathematics-Based Thinking} (MBT),~\textit{Metacognitive Reflection} (MER), and~\textit{Computational Thinking} (CT)~\cite{van2010documenting, manogue2010upper, sayre2015brief, laakso2014promoting, kimbell2011rethinking, nussbaum_dt_failed, rowe1991design, razzouk2012design, ravi_prper_2024, chabay2015matter, boaler2013ability, daly2019mathematical, shute2017demystifying, lopez2023promoting, brown1987metacognition, schraw1994assessing, hamerski_CT_2022, wing2006computational, berland2013student}. Importantly, we acknowledge that WoT frameworks are inherently context-dependent. Even within similar instructional setting, educators may emphasize different aspects based on local constraints, instructional goals, and evolving student needs. This flexibility, which may sometimes be critiqued, is essential in respecting the complexity of educational environments, which are subject to continuous change.

Additionally, the elements within a WoT framework are not mutually exclusive. Overlaps between different~\textit{Ways of Thinking} are not only inevitable but should be embraced. Recognizing these intersections allows educators to align more closely with the way students naturally integrate multiple modes of reasoning and problem-solving~\cite{ravi_SWoT_01_2024, ravi_prper_2024}.

An intentional focus on~\textit{WoT4EDP} can help teachers, researchers, and students by offering a structured yet adaptable framework to support and analyze thinking in engineering design-based STEM tasks. For educators, it provides guidance in designing activities, scaffolding learning, and developing assessment rubrics that capture the integration of design, science, mathematical, computational, and metacognitive thinking. Researchers can use the framework to examine student thinking across various time scales and data sources, gaining insights into how different students engage with complex problems. For students,~\textit{WoT4EDP} fosters deeper engagement, empowering them to direct their own learning by making interdisciplinary connections, engaging in iterative problem-solving, and reflecting on their learning within STEM contexts.~\cite[p.16]{ravi_SWoT_01_2024}.

\subsection{Qualitative Methodology in PER - A Rich Tradition}\label{sec:III.C}

In his compelling, provocative, and incisive 2006 reflections on the state of qualitative research, Amos Hatch, former editor of Qualitative Studies in Education, while advocating for rigorously designed qualitative studies, voiced his concern that:
\begin{quote}
\textit{...qualitative academics and their students seem to be producing fewer data‐based studies at a time when high‐quality qualitative work is at a premium.}~\cite[p.406]{Hatch01072006}
\end{quote}
Shortly after, in 2009, Otero and Harlow~\cite{otero2009getting} provided a foundational guide to qualitative research within the specific context of physics education research (PER). This work served as a crucial resource, offering valuable insights that significantly informed the lead author's approach in all his studies~\cite{subramaniam2023narst, ravi_perc_2023, ravi_perc_2024, ravi_prper_2024, ravi_SWoT_01_2024, ravi_aapt_2024}. Their efforts to instill rigor within PER are reflected in their discussions on data collection, data analysis, the role of theory, among other aspects. 

While numerous studies contribute to the methodological aspects of qualitative research, our exploration led us to a select set of papers within PER that inspired our approach, though this is not an exhaustive list. Hammer explored~\textit{productive student inquiry} in a classroom setting~\cite{hammer1995student}. Henderson {\em et al.}~\cite{henderson_faculty_belief_2007} detail their data collection and analysis techniques using \textit{concept maps}. Kuo {\em et al.}~\cite{kuo2013students} conducted in-depth interview analysis to examine mathematical reasoning in physics, while Redish and Kuo~\cite{redish2015language} used a case study approach to explore how students make sense of mathematical equations in physics. Scherr’s work on~\textit{gesture analysis}~\cite{rachel_scherr_2008} offers a valuable approach to understanding student thinking, with potential applications in our future studies. Given our interest in student metacognition, we also resonate with Sayre and Irving's~\textit{brief, embedded, spontaneous metacognitive (BESM) talk}~\cite{sayre2015brief}. 

Given the rich tradition of qualitative research in PER, there was further amplification through the Physical Review Physics Education Research 2021 \textit{Call for Papers} on Qualitative Methods. However it notes that, while qualitative research is common in PER,
\begin{quote}
~\textit{...the field still lacks a broader use and discussion on many of its current and emergent practices,} 
\end{quote}
further adding,
\begin{quote}
~\textit{questions frequently arise about the validity and reliability of qualitative studies, with both researchers and audiences seeking to ensure qualitative studies are robust.}~\cite{call_for_papers_physrev_2021}. 
\end{quote}

The qualitative research community within PER produced a spirited response to the above call through articles that delve into methodological aspects. In this context, McPadden {\em et al.}~\cite{mcpadden_2023} notably include their research team’s~\textit{positionality} as a validity measure, a practice echoed by Barth-Cohen {\em et al.}~\cite{research_design_cohen_2023}, who advocate for explicit \textit{positionality statements} (see Section~\ref{sec:VII}). They emphasize the importance of \textit{transparency} in ensuring rigor, a stance also reflected in Guisasola {\em et al.}~\cite{phen_graphy_guisanola_2023}. Similarly, Watson and Thomas delve into aspects of ~\textit{credibility}, ~\textit{rigor}, and ~\textit{quality} in qualitative research. Looking ahead, we found Tschisgale {\em et al.}~'s discussion on integrating artificial intelligence into qualitative analysis methodologically relevant, as it addresses not only validity but also ~\textit{scalability}—an aspect we aim to explore in our future work~\cite{Ai_qlr_tschisgale_2023}.

Building on this ongoing discourse, our study engages with these methodological considerations while drawing from established qualitative traditions in PER. We incorporate aspects of reflexivity, transparency, and rigor in our approach, aligning with recent discussions on ensuring the robustness of qualitative research. In the following sections, we outline the specific methodological choices that shape our study.

\section{Methods}\label{sec:IV}

\subsection{Context}\label{sec:IV.A}

This study takes place within a large-enrollment, first-semester calculus-based physics course at a large Midwestern land-grant university in the U.S. The course typically enrolls around 1400 students in the fall semester and 1600 students in the spring semester, with over 80\% aspiring to become engineers and a small fraction pursuing science majors. Adopting a principle-based approach~\cite{chabay2015matter}, the curriculum is divided into three units focusing on fundamental physics principles: momentum, energy, and angular momentum. The reader may see our recent study for more details~\cite{ravi_prper_2024}. The weekly schedule includes two 50-minute lectures, one 110-minute laboratory session, and a 50-minute recitation for problem-solving. Engineering Design (ED) was integrated into the laboratory component in 2019 as part of ongoing reforms.

Table~\ref{tab:sp_23_lab_schedule} outlines the 14-week laboratory schedule for spring 2023. During weeks 01-06, student groups of two or three conducted labs related to the momentum principle, alongside an instructor-assigned ED challenge. Labs focused on the energy principle were completed in weeks 07-09, followed by those on angular momentum in weeks 10-14, during which student groups worked on a student-generated ED challenge.

\begin{table}[!htbp]
\captionsetup{justification=raggedright,singlelinecheck=false} 
\caption{Schedule of laboratory sessions in Spring 2023. Data for this study comes from a lab section with 14 teams, submitted in Week 14. Also see Table~\ref{tab:ED_schedule}. (MP -- Momentum principle; EP -- Energy Principle; AMP -- Angular Momentum Principle)}
\begin{ruledtabular}
\begin{tabular}{p{0.12\linewidth} p{0.4\linewidth} p{0.35\linewidth}} 
\textbf{Weeks} & \textbf{Hands-on lab activities based on} & \textbf{ED problem} \\ 
\hline
01 - 06 & MP & Instructor-assigned \\
07 - 09 & EP & - \\
10 - 14 & MP, EP, AMP & Student-generated \\
\end{tabular}
\label{tab:sp_23_lab_schedule}
\end{ruledtabular}
\end{table}

Python-based computational tasks were incorporated into the physics laboratory in a phased manner, starting in week 01. Students created their lab reports using~\textit{Jupyter Notebooks}~\cite{jupyter} in~\textit{Google Colaboratory}~\cite{googlecolab}. These Notebooks, accessible at no cost with a Google account, combine executable code and rich text, which facilitates running Python scripts, typing equations in LaTeX, uploading images and CSV files, and conducting data analysis without downloads or installations. Students did not need prior programming experience, as the computational tasks focused on applying physics concepts by modifying existing Python code related to the laboratory's physical principles. Carefully designed scaffolds supported students in learning programming in the context of their experiments, which included hands-on inquiry-based activities using PASCO equipment~\cite{pasco_me5300} and software sensors, with data collection and analysis performed using PASCO Capstone™ software~\cite{pasco_capstone}.

\subsection{Student-generated ED problem}\label{sec:IV.B}
From weeks 10 to 14, the student groups identified a real-world Engineering Design (ED) problem of their choice and worked to develop a feasible solution. This process combined weekly hands-on lab experiments with one third of the class time dedicated to the ED task, loosely guided by the ED process model~\cite{ravi_prper_2024}. Outside the laboratory, multi-week activities were supported by scaffolds, including lectures, recitations, and problem solving. Student groups framed their ED problems by identifying clients and stakeholders, stating metrics, outlining criteria and constraints, and making trade-offs. Teaching assistants (TAs) facilitated students' progress toward solutions without influencing their ED problem choices. The detailed schedule of activities planned for weeks 10 - 14 is provided in Table~\ref{tab:ED_schedule}. 

In week 11, a rubric (see~\hyperref[Appendix D]{Appendix D}) was provided that highlighted key aspects of design, science and mathematics to aid their problem-solving process. The rubric outlined: 
\begin{itemize}
    \item \textbf{Design-based Thinking}: Identify stakeholders; outline criteria, and constraints; explore multiple solutions; iterate solutions while considering trade-offs. 
    \item \textbf{Science-based Thinking}: Apply at least two fundamental principles -— momentum, energy, or angular momentum; discuss assumptions and approximations; define the system and surroundings; select appropriate models (point particle or extended body etc). 
    \item \textbf{Mathematics-based Thinking}: Present numerical calculations; use diverse mathematical representations, such as graphs, tables or equations.
    \item \textbf{Python coding}: Adapt code from previous labs for their projects. (Since many were new to Python, our goal was simply to encourage students to use it as a computational tool.)
\end{itemize}

\renewcommand{\arraystretch}{1}
\begin{table*}[!htbp]
\begin{center}
\captionsetup{justification=raggedright} 
\caption{Schedule of lab activities from weeks 10 - 14.  Approximately half of the class time was dedicated to the ED task during weeks 10–13, while the entire class period was allocated to it in week 14. The data for this study consists of transcripts of audio-recordings of group discussions and written reports from 14 student teams in one lab section, submitted in Week 14.}
\begin{ruledtabular}
\label{tab:ED_schedule} 
\begin{tabular}{p{0.03\linewidth} p{0.70\linewidth} p{0.20\linewidth}}
\textbf{Week} & \textbf{Lab activities (Scaffolds)} & \textbf{Deliverables} \\
\hline
10 & \textbf{Hands-on task:} Elastic and inelastic collisions using carts (momentum and energy principles).~\newline{\textbf{ED-task:} Students receive basic information about the student-generated ED-problem and are instructed to individually engage in problem scoping by brainstorming multiple ideas and narrowing their choice to one problem. Student teams asked to engage in a free-flowing group discussion for about three minutes considering stakeholders, criteria, constraints, and relevant physics principles relevant to their problem.} & Weekly lab report + audio recording of the group discussions. \\
11 & \textbf{Hands-on task:} Determining the moment of inertia of a rolling cylinder (energy and angular momentum principles).~\newline{\textbf{ED-task:} Students receive a rubric (see~\hyperref[Appendix D]{Appendix D}) and a \textit{Jupyter Notebook} template for their ED-problem. An example ED-problem similar to those in the first paper~\cite{ravi_SWoT_01_2024} is provided. Students revisit and refine their problem statement if needed, break it into simpler parts, and begin working on the \textit{Jupyter Notebook}, which is due in Week 14.} & Weekly lab report. \\
12 & \textbf{Hands-on task:} Conservation of angular momentum using a rotating turntable and dumbbells (angular momentum principle).~\newline{\textbf{ED-task:} Students continue working on their ED-problem. Students asked to engage in a free-flowing group discussion focused on their specific problem statement and how they intend to apply physics principles in their solution.} & Weekly lab report + audio recording of group discussions. \\
13 & \textbf{Hands-on task:} Determining the moment of inertia of a system about an axis.~\newline{\textbf{ED-task:} Students continue refining and developing their solutions.} & Weekly lab report. \\
14 & \textbf{ED-task:} Entire class time dedicated to the ED-task.~\newline{Student teams are required to submit a final report detailing the problem statement, engagement in design thinking, science thinking, and mathematical thinking, along with their first attempt, first iteration, and an optional second iteration. Teams asked to engage in a free flowing discussion focusing on describing the problem in simple terms, their iterative progression, multiple refinements, and the application of physics principles in their solution.} & Final report (.ipynb file and the PDF version) + audio recording of the group discussions.   \\
\end{tabular}
\end{ruledtabular}
\end{center}
\vspace{-0.5cm}
\end{table*}

\subsection{Data Collection}\label{sec:IV.C}

\begin{quote} 
\small
\textit{Decisions related to sample size in qualitative research are not straightforward.}~\cite[p.416]{milne2005enhancing}
\end{quote}
The data for this study comes from a Spring 2023 laboratory section for which the lead author was the GTA. This dataset was selected for its alignment with key elements of the~\textit{WoT4EDP} framework—design, science, mathematics, computational, and reflective thinking—though the framework was formalized later. The lead author's preliminary studies~\cite{ravi_perc_2024, ravi_aapt_2024} analyzing a subset of this dataset reinforced its suitability, ensuring an organic, unbiased analysis.

The lab section included 42 students across 14 groups, yielding 14 transcripts (of group discussions) and 14 written reports—a substantial, diverse data source for qualitative analysis of students’ design and reasoning. Including all 14 teams supports our goal of informing future pedagogical interventions. Our approach aligns with Oliver’s emphasis on clearly articulating sampling methods:

\begin{quote}
\textit{Methods for sampling should discuss who was selected, how they were selected, how many were involved, and why this strategy was implemented.}~\cite{oliver2011rigor}
\end{quote}

In week 14, student groups submitted reports summarizing their problem statements, design iterations, and solutions, using~\textit{Jupyter Notebooks}~\cite{jupyter} hosted on Google Colab~\cite{googlecolab}. They submitted both .ipynb~\cite{ipynb_2024} and PDF versions of the reports, which integrated text, images, and Python code.

Additionally, students were prompted to submit audio recordings—captured using their mobile phones—of their discussions in response to a set of prompts. They were asked to describe their problem in simple terms, outline their progress toward a solution through multiple iterations, and explain how they applied one or more of the three principles—momentum, energy, or angular momentum—to their solution. Students were asked to specifically discuss how each principle was used, relating key physics concepts to their design decisions. They were encouraged to engage in~\textit{``free-flowing''}~\cite{wu2009improving, murphy2021teacher} discussions for about five minutes.

\begin{figure}[!htbp]
\caption{Data sources: Transcripts and written reports ensure Data Triangulation. Reports include sub-data such as text, images, and Python code.}
\fbox{\includegraphics[width=0.96\linewidth]{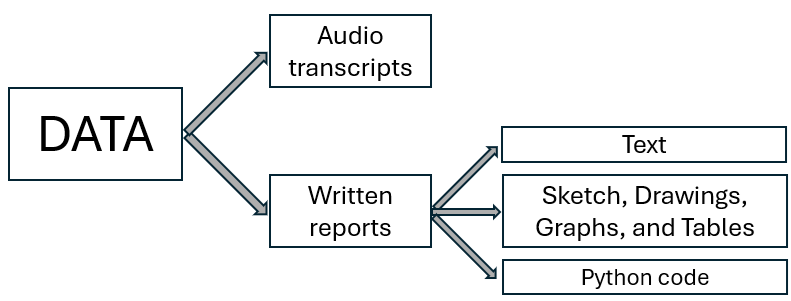}}
\label{fig:data_sources}
\end{figure}

For this study, we analyzed the PDF versions of the final reports and the transcripts of the audio recordings (of all the 14 teams), thereby ensuring data triangulation~\cite[p.37]{brink1987reliability}. See Figure~\ref{fig:data_sources}.

\subsection{Data Analysis}\label{sec:IV.D}

\begin{quote} 
\small
\textit{It is expected that the data analysis be well described, allowing for replication and audit. Questions to be answered include how the data analysis was performed, how the coding was done, and how it is connected to the theoretical framework.}~\cite{oliver2011rigor}
\end{quote}

Student-groups recorded their discussions in the laboratory or corridors, resulting in background noise that made it extremely challenging to uniquely associate conversation segments with individual speakers. Additionally, the final report in week 14 was a collective group assignment. These factors, along with the fact that students worked in groups in all labs, influenced our decision to choose the~\textit{``grain size''} or~\textit{``unit of analysis''}~\cite[p.10]{otero2009getting} to be that of a group. The audio recordings, ranging in duration from 5 to 9 minutes with an average of about 6 minutes, were all manually transcribed by the lead author. The transcript lengths ranged from about 600 to 1200 words each, with the average being about 850. 

The written reports ranged from four to twelve pages, including text, images, graphs, tables, and Python code. The basic requirements included a problem statement, iterations detailing their solutions, and a concluding note (see Section~\ref{sec:V.E(i)}).

Throughout the process, the lead author took detailed handwritten notes, reviewing the data multiple times and documenting reflections and coding decisions. In line with his previous studies~\cite{ravi_prper_2024, subramaniam2023narst, ravi_perc_2023, ravi_perc_2024}, the lead author, as is his wont, adopted a traditional~\textit{pen and paper} approach coupled with~\textit{Word} and \textit{Excel} as his~\textit{``codus operandi''}~\cite[p.26]{saldana2009introduction}, which facilitated direct engagement with the data, allowing for valuable insights to emerge through close interaction with the material. 

Data (transcripts and written reports) from all 14 teams is used for each research question. However, as will be explained in the Findings and Discussion section, the portion of the data selected for each research question varied based on the nature of the question and the suitability of the data. Details are provided in Table~\ref{tab:data_choice}. The data analysis evolved through two stages.

\begin{table}[!htbp]
\captionsetup{justification=raggedright,singlelinecheck=false} 
\caption{Data from all 14 teams were used for each research question. However, the specific portion of data selected for each question varied depending on the nature of the question and the suitability of the data. (RQ -- Research Question; PF -- Problem Framing; DBT -- Design-based Thinking; SBT -- Science-based Thinking; MBT -- Mathematics-based Thinking; MER -- Metacognitive Reflection; CT -- Computational Thinking)}
\begin{ruledtabular}
\begin{tabular}{p{0.04\linewidth} p{0.07\linewidth} p{0.38\linewidth} p{0.25\linewidth}} 
\textbf{RQ} & \textbf{Aspect} & \textbf{Portion of data selected for analysis} & \textbf{Justification provided in} \\ 
\hline
\hyperref[RQ1]{RQ1} & PF & Problem statement in reports & Section~\ref{sec:V.A(ii)} \\
\hyperref[RQ2]{RQ2} & DBT \newline{SBT} \newline{MBT} \newline{MER} & Transcripts and reports \newline{Transcripts and reports} \newline{Transcripts and reports} \newline{Teams' concluding remarks in the reports} & Section~\ref{sec:V.B(i)}  \newline{Section~\ref{sec:V.C(i)} } \newline {Section~\ref{sec:V.D(i)}}  \newline{Section~\ref{sec:V.E(i)}} \\
   
\hyperref[RQ3]{RQ3} & CT & Python code in reports & Section~\ref{sec:V.F(i)}\\
\end{tabular}
\label{tab:data_choice}
\end{ruledtabular}
\end{table}

\subsubsection{Stage 1 coding - Transcripts}\label{sec:IV.D(i)}

\begin{figure}[H]
\caption{Stage 1 coding chart for~\textit{Ways of Thinking}. This chart does not account for the overlap between the various ways of thinking. Additionally, students within a group may repeat ideas, potentially inflating the emphasis on certain elements. In light of these and other limitations discussed in our previous study~\cite{ravi_prper_2024}, we developed a more layered Stage 2 coding approach. Adapted from our recent study~\cite{ravi_perc_2024}.}
\fbox{\includegraphics[width=0.96\linewidth]{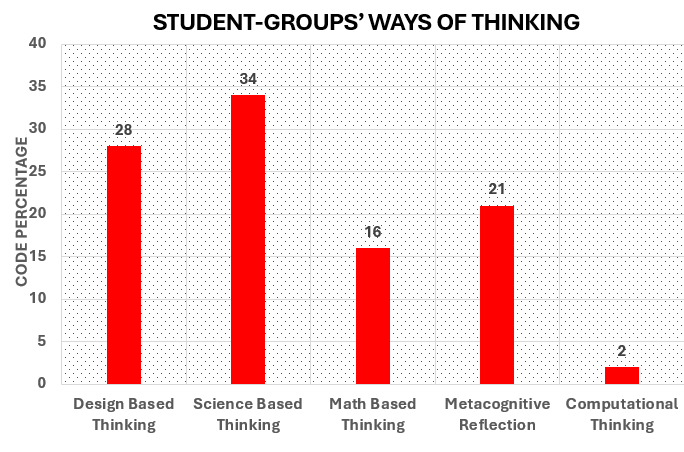}}
\label{fig:wot_disc_chart_1}
\end{figure}

Building on his prior studies~\cite{ravi_perc_2024, ravi_aapt_2024},  the lead author, with assistance from another researcher, coded the transcripts based on five~\textit{Ways of Thinking} in our WoTEDP framework (see Table~\ref{tab:code_descriptors}). Following our prior approach~\cite{ravi_prper_2024}, we used flexible coding units, considering student discussion complexity and group conversation styles~\cite{nihas2020, otero2009getting}. This choice was guided by communication patterns, thematic shifts, speaker transitions, contextual completeness, and our research questions~\cite{ravi_prper_2024}. As Elliot notes~\cite[p.2856]{elliott2018thinking},~\textit{``there is no simple answer''} to the choice of~\textit{``the chunk of data''} to be coded, and~\textit{``it depends on the study and your aims within it''}. Moreover, fixed text lengths might fail to capture the richness and diversity of these discussions, given that some teams engaged in dynamic exchanges in contrast to others that spoke more linearly.  It is also worth noting that multiple codes (code co-occurrence)~\cite{guest2012introduction},~\cite[p.2853]{elliott2018thinking} could be applied to a single segment. To account for the varying transcript lengths and unequal textual segments, we calculated fractional coding frequencies for each team and averaged them across all teams. Given the qualitative nature of the study and our focus on broader trends rather than statistical significance of differences in code percentages,~\textit{error bars have not been included}~\cite{ravi_prper_2024} in Figure~\ref{fig:wot_disc_chart_1}.

\renewcommand{\arraystretch}{1}
\begin{table*}[!htbp]
\begin{center}
\captionsetup{justification=raggedright} 
\caption{The table below shows the~\textit{Ways of Thinking} codes for transcripts, code descriptions, example quotes, and code justification. (DBT--Design-Based Thinking; SBT--Science-Based Thinking; MBT--Mathematics-Based thinking; MER--Metacognitive Reflection; CT - Computational Thinking.)}
\begin{ruledtabular}
\label{tab:code_descriptors} 
\begin{tabular}{p{0.02\linewidth} @{\hspace{0.5cm}} p{0.28\linewidth} p{0.38\linewidth} p{0.22\linewidth}}
    \textbf{Code} & \textbf{Code Description} & \textbf{Example Quote} & \textbf{Code Justification} \\
    \hline
    DBT & State the problem; identify criteria and constraints; brainstorm multiple solutions; iterate, select the best solution; consider design aspects; prototype the solution; communicate.~\cite{engineering_design} & \hl{\textit{``The axe handle has a length of 0.36 m and a one-to-one ratio for the mass of the axe handle to the axe head. The National Building Code regulates that the ceiling needs to be at least 3 m above the highest point the axe can travel for the average customer.''}} (Team-13) & Considering design dimensions and drawing reference to the mandates of the National Building Code (criteria and constraints). \newline{- Also coded for MBT (numerical aspects related to the contraption).}\\ 
    SBT & Identify related physics terms, concepts, or principles; cause and effect; system and surroundings; scale; change and rate of change. \cite{reachoutmichigan_concepts} & \hl{\textit{``Since we had the arrow’s final KE, we could solve for the amount of PE at the start, because the arrow was stationary before it was fired.''}} (Team-05) & Use of terms such as KE (kinetic energy) and PE (potential energy), stating initial velocity, and indirectly implying the use of energy conservation principle.\\ 
    MBT & Mention a formula, equation, or a mathematical concept; refer to a scientific statement in terms of a relation among several variables and constants; proportional reasoning; units analysis; use of explicit equations. & \hl{\textit{``Torque times delta t = [change in] angular momentum.''}} (Team-08) & Direct use of a physics formula and employing the language of mathematics. \newline{- Also coded for SBT (applying physics principles).} \\ 
    MER & Reflect on their design and science ideas and progression towards the solution \cite{vanderbilt_metacognition}. & \hl{\textit{``As far as the angular momentum principle is concerned, we haven’t incorporated it into our solution at the present time although if we were to continue further iterations of the solution perhaps we can look into ideas such as the torque of the wheel to better understand and create a more detailed design.''}} (Team-02) & Thinking about ideas which are yet to be made use of, and which may be put to use at a later stage.\newline{- Also coded for DBT (to improve design) and SBT (referring to physics principles).}\\ 
    CT & Make references to Python codes used in their written reports.~\cite{shettigar2023thinking} & \hl{\textit{``We have the translational KE and I [moment of inertia] again put that as a float so we can use that as numbers rather than a string.''}} (Team-03) & Making a direct reference to their Python code. \newline{- Also coded for SBT (use of physics terms).}\\ 
\end{tabular}
\end{ruledtabular}
\end{center}
\vspace{-0.5cm}
\end{table*}

Having detailed the coding process in this first stage, we emphasize that our findings do not majorly depend on this initial coding. As outlined in our previous study, this approach presents several challenges. First, textual segmentation—while a common method—tends to fragment students’ ideas, undermining our goal of understanding their thinking. Additionally, in a hypothetical extreme case, a team might repeatedly state a particular idea (e.g., reflecting design-based thinking), which would artificially inflate the corresponding code. The lead author carefully considered these limitations and developed a more rigorous coding scheme to address them~\cite{ravi_prper_2024}.

The reason we describe this stage in detail despite not relying on it extensively is methodological. This stage -- a crucial first step in our analysis -- provided important insights that motivated the development of a more layered coding approach. Moreover, this initial method proves inadequate for analyzing written reports, where the data is more complex, integrating text, Python code, figures, graphs, equations, and tables. Furthermore, the diversity in teams' writing (including diagrams and coding) styles added yet another layer to the already intricate and multifaceted nature of the data, which O'Dwyer terms~\textit{``a messy but attractive nuisance''}~\cite{odwyer_qlr_messy_intimate}.  As in our previous study~\cite{ravi_prper_2024}, we conducted a second round of coding, which we describe in the following section.

\subsubsection{Stage 2 coding - TBEER}\label{sec:IV.D(ii)}

To our~\textit{blended} analytical approach augmented by the Gioia~\cite{gioia2013seeking} and MIRACLE~\cite{younas2023proposing} frameworks. We refined and expanded on the multi-layered coding schema from our recent study~\cite{ravi_prper_2024}.~\hyperref[Appendix A]{Appendix A},~\hyperref[Appendix B]{Appendix B}, and~\hyperref[Appendix C]{Appendix C} provide the coding schemata developed separately for the transcripts and written reports. 

To address~\textit{reliability}, paying heed to the caution sounded by Morse {\em et al.}:  
\begin{quote}
\textit{...by focusing on strategies to establish trustworthiness (Guba and Lincoln’s 1981 term for rigor) at the end of the study, rather than focusing on processes of verification during the study, the investigator runs the risk of missing serious threats to the reliability and validity until it is too late to correct them.}~\cite[p.14]{morse2002verification}
\end{quote}
the lead author transitioned to a~\textit{``constructive (during the process)''}~\cite[p.15]{morse2002verification} procedure as opposed to a traditional~\textit{``evaluative (post hoc)''}~\cite[p.15]{morse2002verification} inter rater reliability (IRR) test. Taking guidance from Cascio {\em et al.}~\cite{casio_icc_team_based}, we developed a team-based approach which we term~\textit{Team-Based Efforts to Enhance Reliability} (TBEER). Over a two-month period, two coauthors (Nikhil Borse [NB] and Winter Allen [WA]) assisted in the analysis, meeting with the lead author at least once a week for about an hour. While lead author coded all 14 transcripts and 14 reports, NB coded 7 transcripts and 7 reports, and the remaining transcripts and reports were coded by WA. The lead author, NB, and WA independently coded the transcripts and written reports. They then compared, discussed, and debated their codes to iteratively reach a~\textit{consensus}. Cascio {\em et al.} term this an~\textit{inter-coder consensus} as opposed to \textit{inter coder reliability}~\cite{casio_icc_team_based}, while Linneberg and Korsgaard~\cite[p.268]{linneberg_novice_coding} term it~\textit{inter-coder agreement}. Through regular~\textit{``peer debriefing''}~\cite[p.14]{morse2002verification},~\cite{oliver2011rigor} sessions, the team collaboratively moved from data to findings. The lead author meticulously maintained an~\textit{audit trail} of the team's deliberations through the study~\textit{``to document the course of development of the completed analysis''}~\cite[p.19]{morse2002verification}.

Finally, to further strengthen the~\textit{reliability} of our findings, we engaged an additional researcher from our team, specializing in Python coding, to review our analysis related to computational thinking (see Section~\ref{sec:V.E}). This external review provided an additional layer of verification, ensuring that our interpretations of students' computational engagement were both accurate and well-founded.

As will become evident in the following sections, our findings primarily rely on the second stage of coding, though the first stage played a crucial role in shaping and informing this process. We also draw insights from the first stage where relevant.

\subsubsection{Guiding Methodological Perspectives}\label{sec:IV.D(iii)}

It may not be an exaggeration to say that every qualitative study is unique~\cite{unique_qlr_delmar_2010}. With this in mind, while we draw inspiration from recent studies in PER, we also integrate a range of methodological perspectives to inform our approach.

While our goal is to analyze the data using our~\textit{WoT4EDP} framework as a lens, we also followed the principle of~\textit{``letting the data speak to the researcher''}\cite[p.8]{urquhart2022grounded}. This~\textit{``blended approach''}~\cite[p.264]{linneberg_novice_coding} enabled us to integrate structured analysis with openness to emergent insights, making it well-suited to our study.

Given that our data is rooted in a physics laboratory incorporating ED-based tasks, our analysis and interpretation were deeply informed by our disciplinary understanding of Newtonian Mechanics. Although students occasionally referenced concepts from other areas of physics, their science-based thinking was primarily within the conceptual structure of Newtonian Mechanics~\cite{ravi_prper_2024}.

To be systematic, we made a conscious effort to maintain an~\textit{``audit trail''}~\cite[p.264]{beck1993qualitative},~\cite[p.10]{nazar2023evidence} following Lincoln and Guba's six-aspect guideline summarized by Nazar {\em et al.}, specifically: (a) records of all raw and secondary data, (b) activity logs of discussions with co-researchers, (c) decision logs with hand-written notes and comments, (d) data analysis records and documenting code sheets, and (e) documenting the work of those who influenced the study~\cite[p.10]{nazar2023evidence}. 

Further, we were guided by Tracy's eight~\textit{big-tent} criteria for excellence to ensure quality in qualitative research', namely: (a) worthy topic, (b) rich rigor, (c) sincerity, (d) credibility, (e) resonance, (f) significant contribution, (g) ethics, and (h) meaningful coherence~\cite[p.840]{tracy2010qualitative}.

Saldana, Guest {\em et al.}, and Otero and Harlow provided essential guidance as we engaged in traditional coding, which remains a central component of our studies~\cite{saldana2009introduction, guest2012introduction, otero2009getting}. Structure in our coding process developed over a period of time, although in the initial stages coding was largely~\textit{``instinctual''} and based on~\textit{trial and error''}~\cite[p.2850]{elliott2018thinking}, and of course~\textit{``on the job''}~\cite[p.475]{seale2000quality} learning. Although our coding began with the textual segmentation of the transcripts, we later transitioned to a more holistic analysis~\cite{ravi_prper_2024}. Our coding -- (or, do we say~\textit{decoding}?~\cite{decode_cooper_2009},~\cite[p.5]{saldana2009introduction})-- process evolved over several months and involved multiple researchers who, as graduate teaching assistants, were familiar with the data and its import. 

In addressing qualitative rigor, the qualitative literature is replete with a~\textit{``plethora of terms and criteria''}~\cite{morse2002verification}, including~\textit{reliability},~\textit{trustworthiness},~\textit{credibility},~\textit{dependability},~\textit{fittingness},~\textit{auditability}~\cite{brink1987reliability, beck1993qualitative} to list a few. Notably, some scholars even question whether the traditional~\textit{inter-rater reliability} (IRR) test truly fulfills its intended purpose~\cite{mcdonald2019reliability, morse1999myth}. In their insightful essay, McDonald {\em et al.} raise questions that may challenge conventional wisdom. Their pithy reflections, such as~\textit{``To IRR or not to IRR''} and~\textit{``To IRR is human''}~\cite[p.72.1, 72.5]{mcdonald2019reliability}, offered us valuable food for thought and deepened our own reflections on the subject. 

Addressing the perceived subjectivity often associated with qualitative research, Peshkin directly engages with this concern in his insightful essay, citing Cheater’s caution:
\begin{quote}
\textit{We cannot rid ourselves of this subjectivity, nor should we wish to; but we ought, perhaps, to pay it very much more attention...}~\cite{peshkin1988search}
\end{quote}

Following this guidance to pay~\textit{``very much more attention''} and drawing on Morse et al.~\cite{morse2002verification}, we developed a process called~\textit{Team-Based Efforts to Enhance Reliability} (TBEER), detailed in Section~\ref{sec:IV.D(ii)}.

Tuval-Mashiach~\cite{tuval2017raising}, while encouraging the researcher~\textit{``to craft his or her own method''},  advocates~\textit{``raising the curtain''} to promote greater transparency in qualitative research. Although we provide complete coding details, following Guest {\em et al.}'s caution that~\textit{``[code] frequency alone cannot tell you the importance of a given theme''}~\cite[ch.6]{guest2012introduction}, our findings are not solely based on code frequencies. Instead, we employ thematic analysis and thick description to enhance the depth and nuance of our reporting. 

Our multi-layered coding was inspired by the Gioia Framework~\cite{gioia2013seeking, ravi_prper_2024}, while the MIRACLE framework~\cite{younas2023proposing} guided the presentation of our findings through~\textit{Thick Description}~\cite[p.26]{stahl2020expanding},~\cite[p.38]{brink1987reliability}, allowing~\textit{``the reader [to] enter the research context''}~\cite[p.12]{peel2020beginner}. 
Elliot~\cite{elliott2018thinking}, Braun, and Clarke~\cite{elliott2018thinking, braun2006using} guided our thematic analysis. We consider a theme to be~\textit{``something important the data in relation to the research question...''}~\cite[p.82]{braun2006using}. In our thematic analysis~\cite{braun2006using},~\cite[p.8]{peel2020beginner}, we integrate~\textit{Thick Description} with our interpretations~\cite{ravi_prper_2024}. Ponterotto, while acknowledging the~\textit{``confusion''} surrounding what constitutes a~\textit{thick description}, offers some operational guidelines, which we found useful in guiding our approach. In simple terms, it can be understood as providing rich details of the findings and discussions (in addition to the methods) so as to offer the reader a~\textit{``sense of verisimilitude''}~\cite{ponterotto2006brief}. We did not feel compelled to strictly adhere to any single approach. Reay {\em et al.} emphatically affirm that~\textit{``one size does not fit all!''}, and add cautionary note that blind adherence to templates may lead to a state where~\textit{``rigor can become rigor mortis''}~\cite[p.202]{reay2019presenting}.

The coding charts enhance~\textit{dependability}~\cite{stahl2020expanding}, while our data-driven interpretations provide deeper insight. Carefully selected participant quotes reinforce trustworthiness and connect readers to the data. Eldh {\em et al.}\cite[p.4]{eldh2020quotations} highlight the need for effective use of quotations, clear attribution, and consistency between data and findings. Example quotes, including figures, were carefully chosen to illustrate our findings and we took care to have all teams and all types of data fairly represented, adhering to Eldh {\em et al.}:
\begin{quote}
~\textit{...quotations preferably apply for illustrating the analysis process and/or findings, while the idea that quotations can be employed to validate findings has
limited support.}~\cite{eldh2020quotations} 
\end{quote}

Adhering to Wa-Mbaleka's guidance that qualitative research~\textit{``may be messy, [but] we need to explain what we did, how we did it, and why we did it''}\cite[p.40]{wa2020researcher}, we strove to enhance~\textit{transparency}~\cite{tuval2017raising} and~\textit{trustworthiness}~\cite{jacobs2019transparency, moravcsik2014transparency} of our findings. To enhance transparency, we integrate our methodology throughout the manuscript. While broad details are introduced is this section, for clearer communication, finer aspects are presented in the next section on Findings \&\ Discussions, which is directly related to the research questions. 

Additionally, some aspects of methodology are revisited in the \textit{reflexivity statement} in Section~\ref{sec:VII}. The reader may also find it useful to refer to our earlier studies~\cite{ravi_prper_2024, ravi_SWoT_01_2024} or contact us directly.

\section{Findings \&\ Discussion}\label{sec:V}

In this section, we present our findings on how the 14 student teams engaged with the five elements of the~\textit{WoT4EDP} framework. Section~\ref{sec:V.A} addresses our first research question~\hyperref[RQ1]{RQ1}. Sections~\ref{sec:V.B},~\ref{sec:V.C},~\ref{sec:V.D}, and~\ref{sec:V.E} explore our second research question~\hyperref[RQ2]{RQ2}. Section~\ref{sec:V.F} examines computational thinking in relation to our third research question~\hyperref[RQ3]{RQ3}. 

In our pursuit of methodological transparency, specific to each research question, we amplify the analytical details in Section~\ref{sec:V.A(ii)}, Section~\ref{sec:V.B(i)}, Section~\ref{sec:V.C(i)}, Section~\ref{sec:V.D(i)}, Section~\ref{sec:V.E(i)}, and Section~\ref{sec:V.F(i)}. Although this might appear to be an overlap with the Methods section, this is in line with the recommendation of Devetak {\em et al.} that the strength of good qualitative research lies in offering~\textit{``sufficient details for the reader to grasp the idiosyncrasies of the situation''}~\cite[p.83]{devetak2010role}. 

To further orient the reader, we highlight a key consideration in addressing the research questions: uncovering patterns that may inform future pedagogical interventions. While we remain mindful of the limits of generalization, we aim to identify insights that hold local~\textit{transferability} within our educational setting~\cite{maxwell2014generalization, hallberg2013quality}. To this end, each subsection concludes with a `Moving Forward' note, offering suggestions for improvement in our next iteration of pedagogical intervention.

In presenting our findings, we maintain the qualitative focus by following the guidance of Guest {\em et al.} in describing the prevalence of a code among the 14 teams:
\begin{quote}
\textit{Some researchers advocate using quantitative descriptors only (all, many, most, some, few), and it is possible that you might even define what these terms mean at the outset of your paper.}~\cite[p.156]{guest2012introduction}.
\end{quote}
As a rule of thumb, we use terms such as `a majority' or `a significant number' or `most' to refer to more than 8 teams, `about half' for 6–8 teams, and a `smaller fraction' or `only a few' for fewer than 6 teams. In our view, precise numbers may not be particularly relevant in our qualitative study, though the coding sheets (see~\hyperref[Appendix A]{Appendix A},~\hyperref[Appendix B]{Appendix B}, and~\hyperref[Appendix C]{Appendix C}) do reveal them.

We reiterate that~\textit{Ways of Thinking} are not isolated elements. While they are interconnected, in this study, we focus on discussing these elements separately, though we occasionally highlight their interconnections. However, we will explore this aspect in depth in the third part of this series, focusing on \textit{Integrative Thinking}. Readers may find it useful to refer to our recent study, where we examined \textit{design-science connections}~\cite{ravi_prper_2024} in students' thinking as they navigated an instructor-assigned ED task.

\subsection{ED-problem Framing}\label{sec:V.A}

In this section, before analyzing students' solutions, we first examine aspects of their problem framing, given that the problem itself was framed by the students. To emphasize the significance of problem framing, we provide an introductory discussion in Section~\ref{sec:V.A(i)}. Section~\ref{sec:V.A(ii)} details our approach to examining the data, Section~\ref{sec:V.A(iii)} outlines key findings, and Section~\ref{sec:V.A(iv)} offers recommendations to support students in refining their problem framing.

\subsubsection{Problem Framing - A key aspect of design}\label{sec:V.A(i)}

Problem framing is a crucial part of the design process, significantly shaping the final outcome. However, research suggests that it remains an understudied topic~\cite[p.1]{dinar2011towards}. Cross and Cross~\cite{cross1998expertise} highlight the importance of studying expert designers’ approaches to systematically develop problem-framing skills, which is essential for transitioning from novice to expert designers. Their examples involving the application of physics concepts within engineering design were particularly relevant to our work. As they note:~\textit{``there remain considerable difficulties about explaining and learning how to frame problems in creative ways''}~\cite[p.147]{cross1998expertise}. 

While giving students the agency to frame and solve their own problems is valuable, it comes with challenges. Most students are accustomed to solving \textit{well-defined} textbook problems and rarely have opportunities to frame their own problems~\cite{well_ill_simon_1973, well_strcutured_milbourne_2018}. To address this lacuna, we began the semester with an engineering design (ED) problem~\cite{ravi_prper_2024} that we had framed (see Table~\ref{tab:sp_23_lab_schedule}). Later in the semester, we introduced the student-generated ED problem task (see Tables~\ref{tab:sp_23_lab_schedule} and~\ref{tab:ED_schedule}). In week 10, student teams were asked to:
\begin{quote}
\textit{identify a real-world problem and propose a feasible solution aligned with Momentum, Energy, and Angular Momentum principles while considering stakeholders, criteria, and constraints.}
\end{quote}

Students were guided to progress towards their solutions from weeks 10 - 14. Students were encouraged to engage with TAs for guidance, while TAs were instructed to remain flexible and supportive, even when students set ambitious goals (see Section~\ref{sec:II}). 

Our approach of guiding students toward framing a problem statement as a key competency aligns with the perspective that \textit{``developing competencies to solve ill-structured problems in teams is essential for engineering students''}~\cite[p.460]{dringenberg2018experiences}.

\subsubsection{Problem Framing -- Examining the data}\label{sec:V.A(ii)}

\begin{figure}[H]
\caption{One of the iterations of the coding sheet for analyzing students' design-based thinking (inclusive of framing the problem statement) is presented below. The final results may be seen in~\hyperref[Appendix A]{Appendix A}.}
\fbox{\includegraphics[width=0.96\linewidth]{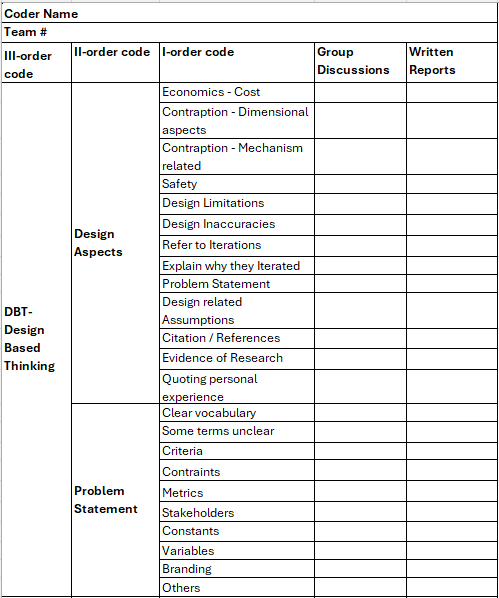}}
\label{fig:code_sheet_dbt_blank}
\end{figure}

As shown in Figure~\ref{fig:code_sheet_dbt_blank}, we categorized the problem statement under design-based thinking. This classification aligns with the perspective that design thinking is a problem-solving approach that integrates creative thinking to drive innovation. A fundamental aspect of design thinking is \textit{problem framing}, which shapes the direction of the design process~\cite[p.496]{prob_framing_pham}. It may be noted that since students are expected to apply physics concepts, an overlap with science-based thinking is not only inevitable, but indeed welcome.  

Analyzing the students' problem statements (see Table~\ref{tab:ED problems list}) presented two main challenges. First, each team worked on a unique problem, and second, some teams presented their problem in scattered paragraphs that sometimes blended into their solutions. Accordingly, our analytical approach showed flexibility to how students presented their problem statements. 

An iteration of the coding sheet used by our research team may be seen in Figure~\ref{fig:code_sheet_dbt_blank}. The coders would take notes in the sheet independent of each other, discuss, and arrive at a consensus. The process was repeated for each team for both transcripts and written reports, not just for this section but right through the study.   
By compiling the data in all the coding sheets, we developed a final consolidated coding table, which can be found in~\hyperref[Appendix A]{Appendix A}. As an interesting coincidence, our coding scheme aligns with certain aspects of problem framing as outlined by Pahl et al.~\cite{pahl2013engineering}. We want to emphasize that the aspects we focused on are specific to our interests as educators and the local context. We encourage educators to remain flexible and adapt to their own unique contexts.

\renewcommand{\arraystretch}{1}
\begin{table*}[!htbp]
\begin{center}
\captionsetup{justification=raggedright} 
\captionsetup{width=\textwidth}
\caption{Table presents a broad overview of the student-generated ED problems taken for this study.}
\resizebox{\textwidth}{!}{%
\begin{ruledtabular}
\label{tab:ED problems list} 
\begin{tabular} {p{0.07\linewidth} p{0.20\linewidth}p{0.70\linewidth}}
\textbf{Team \#\ } & \textbf{Topic} & \textbf{Problem objectives}\\
\hline

01. & Baseball: \newline{Home-Run Conditions} & To explore the optimal conditions for a baseball player to hit a home run by applying physics concepts to calculate the ball's trajectory, accounting for factors like pitch speed, spin, timing, position, and technique from the batter's perspective. \\
02. & Self-Driven Golf Carts & To determine the minimum safe travel time and engine work needed for self-driving golf carts over various distances, using physics principles to ensure efficiency and safety. \\
03. & Mortar Launch: \newline{Calculation System} & To calculate a mortar's range using launch angle and translational kinetic energy, providing the military and soldiers with data for effective field use. \\
04. & Catapult Launch: \newline{Range vs. Angle} & To determine the optimal launch angle for a catapult projectile to maximize its range, addressing logistical and safety concerns in battlefield use. \\
05. & Hookean Bow: \newline{Strength of Drawstring} & To determine the necessary strength of a bow's drawstring, modeled on Hooke's law, to achieve a shot of  a prescribed range, angle, and launch speed, ensuring it meets professional standards for performance and precision. \\
06. & Racing Cars:\newline{Max. speed on a Racing Track} & To calculate the maximum speed a racing car can achieve around corners with prescribed radii, considering grip dependent solely on the car's mass, and to determine the corresponding kinetic energies at these speeds.\\
07. & Mail:\newline{Delivery by Drones} & To develop a program to optimize a delivery drones' efficiency and safety, ensuring timely mail delivery within prescribed  constraints of travel height and load limit.\\
08. & Carpentry Tool:\newline{Hammer Design} & To determine the weight and dimensions of a hammer used by carpenters based on usage needs, efficiency, and cost preferences. \\
09. & Duncan's Toy Chest:\newline{Toy Catapult} & To calculate the distance a foam projectile will travel after being launched by a toy catapult, ensuring the design meets safety standards as required by a toy making company. \\
10. & Bulletproof Vest: \newline{Stopping Distance} & To calculate the force required to stop a bullet of a given mass, shot from a specified distance at a prescribed speed, by a bulletproof vest of a given thickness.\\
11. & Fungee Jumping: \newline{Maximum Load} & To determine the maximum load that can be safely attached to a bungee cord of a given spring constant and relaxed length, to ensure that a jumper stops above a specified safe distance from the ground. \\
12. & Roller Skates Design & To determine the final speed of the skater and the angular velocity of the wheels as they skate down a slope of given initial height,given the skater's mass and wheel dimensions.  \\
13. & Axe Throwing & To determine the maximum height and distance an axe of a given mass travels when thrown at a prescribed angle, considering user-applied force, throw time, and average height, to guide building facility design.\\
14. & Yo-Yo speed & To determine which of the three given Yo-Yos will have the fastest final angular speed upon falling, considering their dimensions and masses.  \\
\end{tabular}
\end{ruledtabular}
}
\end{center}
\end{table*}

Students began working on the~\textit{Jupyter Notebooks} for the ED problem in week 10, but our analysis focused solely on the final week's PDF report. We relied on these reports rather than group discussions, as most teams prioritized solutions over revisiting their problem statements in their discussions (see Table~\ref{tab:data_choice}). This analysis examines the final product of problem framing, not its evolution over time. For coding results, refer to~\hyperref[Appendix A]{Appendix A}. Figures~\ref{fig:ED_ps_05} and~\ref{fig:ED_ps_09} present the problem statements from two teams.

\begin{figure}[tb]
\centering
\begin{tcolorbox}
\justify{We are an Engineering Firm that focuses mainly on quality and testing engineering, working with a lot of companies to make sure their designs are good enough to fit whatever specifications they want. We have been reached out by a new professional Bow-making company, named Hookean
Bows, that has some questions about how they should design their first bow and arrow. They have a bow that models a Hookean Spring when the drawstring is pulled back, and they have their own arrow and target for testing. Hookean Bows wants to know how strong the drawstring needs to be to achieve a standard Compound bow shot. Standard bow shots vary, but the standard shot of an archer involves them pulling back the bow on a standard range. The specifications are a range of 50 m from the bow, with a target at the same elevation of the arrow when it leaves the bow (Olympics).
The bow should have an angle of elevation of 6 degrees above the horizontal, and the bow needs to be strong enough to have the arrow hit the target at 61 meters per second, the average speed of an arrow fired (Archery Tips). This is horizontal speed only. The arrow itself is 500 g. Due to it's very low angle of elevation, we can assume the vertical speed of the arrow to be very much less than the horizontal speed, approximating to zero. Since it's a standard shot, there will be no wind and there
will be negligible air resistance, thanks to the arrows' construction, so the only force besides the bow is gravity. The bow itself has the property of being the same as a hookean spring, with it's normal L being 0.2 m `away' from the start of the bow, and it needs to be strong enough to hit the desired shot with a stretch of 0.75 m, the standard bow pull of archers.}   
\end{tcolorbox}
\caption{Example problem statement from Team-05. The team identified stakeholders (engineering firm, companies), provided the reasoning behind their problem choice, outlined criteria (range and target location) and constraints (specific angle of elevation), modeled the spring to obey Hooke's law, provided metrics (numerical factors relevant to their problem), stated assumptions (negligible air resistance), and even designated a name for their company. See~\hyperref[Appendix A]{Appendix A}.}
\label{fig:ED_ps_05}
\end{figure}

\begin{figure}[tb]
\centering
\begin{tcolorbox}
\justify{With the world of toys becoming more and more competitive, companies like Duncan's Toy Chest are doing whatever they can to make better and better toys. Duncan's Toy Chest has commissioned our firm to create a toy catapult that is better than any other before it. This catapult must be mostly plastic with foam pieces in order to minimize any possibility of harm to the user. This includes a soft projectile and a covered spring to initiate the launching process. While it would be simple to make a trebuchet like toy that launches the projectile by using a mass on the end instead of the spring, Duncan's Toy Chest has requested that our group find the final distance of the bean bag from the point of launch, after it is launched. The reasoning for the maximum height to be found, is so that the company can ensure parents that the toy is safe to use indoors and outdoors.}   
\end{tcolorbox}
\caption{Example problem statement from Team-09. The team identified the stakeholders (the company, users (children), parents), provided a reasoning behind their problem choice, stated criteria (material choice, safety) and constraints (safety), did not explicitly mention any numerical factors, and even designated a name for their company. See~\hyperref[Appendix A]{Appendix A}.}
\label{fig:ED_ps_09}
\end{figure}
\subsubsection{Problem Framing -- What the data reveals}\label{sec:V.A(iii)}

A notable observation was that no two teams selected the same problem (see Table~\ref{tab:ED problems list}). This diversity in problem selection suggests that students brought their own creative perspectives, choosing problems that aligned with their unique interests. The variety of problems also indicates a genuine effort to explore different real-world challenges rather than defaulting to a standard or predefined template. Moreover, the uniqueness of each problem statement (examples in Figures~\ref{fig:ED_ps_05} and~\ref{fig:ED_ps_09}) highlights the flexibility of the open-ended design task in fostering student ownership and agency in problem framing.

Almost all teams identified stakeholders in their problem statements, with some teams stating it explicitly. As an example, Team-01 stated: 
\begin{quote}
\hl{\textit{``we are considering our stakeholders to be the players, specifically the batter and pitcher, baseball teams, baseball coaches and fans.''}}
\end{quote}

All teams revealed their problem motivation by establishing relevance through a social context, often making for an engaging and compelling read. As an example, Team-06 started by stating:
\begin{quote}
\hl{\textit{``In motorsport, speed and grip are vital components of a race, both of which are dependent on the
car. Teams typically want to maximize grip in order to achieve the highest possible speed their
car can drive around the track.''}}
\end{quote}

Team-08, in addition to providing context in their problem statement, revisited these ideas later in their report by sharing one particular engineering-based fact which we found to be particularly revealing and worth recounting:
\begin{quote}
\hl{\textit{``While some people might as why do we even still use nails when screws and electric drills have existed for so long, it is because nails in certain instances like in the
construction of houses, allow for slight movement in cases of small earthquakes, where screws do not.''}}
\end{quote}
The lead author was unable to verify the accuracy of this specific claim but found it thought-provoking. This insight was new to the lead author, highlighting the valuable knowledge our students bring to the table. It is clear there is much we can learn from them.

Nearly all teams incorporated numerical parameters (metrics) in their problem statements in varying degrees, with more details emerging later in their reports. Most teams explicitly stated criteria and constraints in their problem statements (for eg., see Figure~\ref{fig:ED_ps_05}), with some adding further details later (for eg., see Figure~\ref{fig:ED_ps_09}).

An interesting trend emerged: about half the teams assigned creative branding or marketing names to their designs, reflecting an entrepreneurial mindset~\cite{crilly_design_2024} and engagement beyond mere technical problem-solving. Examples include Fungee Jumping (Team-11's bungee jumping design), Hookean Bow (Team-05's bow and arrow), and Duncan's Toy Chest (Team-09's bean bag toy). These names showcased students' creativity and marketing instincts, highlighting an aspect of design thinking that, while valuable, may not be typically prioritized in a physics classroom. See Table~\ref{tab:ED problems list}.

Importantly, all teams connected their problem, either directly or indirectly, to physics and mathematics concepts. While constructing a physical model was impractical within our course constraints, students navigated this by either focusing on engineering design process models or conceptualizing contraptions, or a blend of these two. See~\hyperref[Appendix A]{Appendix A}.

We also observed that while most teams addressed the notion of `system and surroundings' in their solutions, a couple of teams chose to incorporate within their problem statement. In our view, this is a valuable practice in a physics course, as it facilitates the application of fundamental principles such as momentum, energy, and angular momentum. For instance, Team-11, when discussing the person experiencing the bungee jump, stated:

\begin{quote}
\hl{\textit{``Energy principle is useful because we are able to set the initial energy of the system equal to the final energy of the system.''}}
\end{quote}
This example also serves to illustrate the variety of ways students chose to frame their problem statement.

About half of the teams explicitly referenced assumptions and approximations in their problem statements. This may reflect an awareness of design complexities or an implicit plan to address these aspects later. A challenge we encountered was distinguishing between assumptions and approximations, as students often used these terms interchangeably. We adopted a flexible approach to accommodate the different ways in which teams chose to communicate. For example, Team-05 stated:
\begin{quote}
\hl{~\textit{``Due to it's [arrow's] very low angle of elevation, we can assume the vertical speed of the arrow to be very much less than the horizontal speed, approximating to zero. Since it's a standard shot, there will be no wind and there will be negligible air resistance''}}. 
\end{quote}
Stating that the arrow's initial vertical speed is close to zero can be viewed as either an approximation (estimating it as zero for convenience) or an assumption (choosing to ignore a minor detail). 

\subsubsection{Problem Framing -- Moving forward}\label{sec:V.A(iv)}

In physics classrooms, use of terms such as design, stakeholders, criteria, constraints, and metrics is generally less common. By providing examples, students may need to be guided in addressing these aspects in the problem statement. Students may be encouraged to state any assumptions or approximations upfront, acknowledging that these may evolve as they progress through their solution. They may be asked to consider real-world factors such as cost and safety, while maintaining flexibility to refine their problem statement as their understanding deepens. 

In engineering design-based activities, the role of iterative thinking may need to be emphasized. Following a systematic approach to \textit{brainstorming}~\cite{kaufman2002getting} which we had outlined in our recent paper~\cite{ravi_prper_2024}, may aid in guiding students to frame their problem statement. Valkenburg and Dorst~\cite[p.254]{valkenburg1998reflective} emphasize that reflective practices can help rethink and reframe problems, leading to better solutions.

It is also important for students to select problems that not only align with their personal interests but are reasonably simple so that meaningful progress is possible given limited time and resources. Actively seeking input from TAs—can enhance problem framing and engagement. 

\subsection{Design-Based Thinking (DBT)}\label{sec:V.B}

In this section, we explore how design-based thinking (DBT) emerges in students' work. Section~\ref{sec:V.B(i)} describes our approach to examining the data, Section~\ref{sec:V.B(ii)} presents key findings, and Section~\ref{sec:V.B(iii)} provides guidance for strengthening students' design thinking skills. (See Table~\ref{tab:wot_description}). 

\subsubsection{DBT -- Examining the data}\label{sec:V.B(i)}
The lead author conducted a preliminary study in Fall 2022~\cite{subramaniam2023narst}, which helped identify common gaps in how students navigate ED-based tasks. Insights from that study informed the development of our rubric (see~\hyperref[Appendix D]{Appendix D}) for the ED task in Spring 2023. To support design-based thinking, we encouraged students to consider key aspects such as stakeholders, criteria and constraints, iterative refinement of their solutions, and any trade-offs (factors that weigh against each other while deciding the best possible solutions) they made. See Table~\ref{tab:ED_schedule}.

For the current study, while our coding of DBT was guided by our recent publications~\cite{ravi_prper_2024, ravi_SWoT_01_2024} and our~\textit{WoT4EDP} framework, we adopted a blended approach by remaining open to what the data might reveal. Figure~\ref{fig:code_sheet_dbt_blank} presents an iteration of the code sheet used by the coders, while Figure~\ref{fig:DBT_code_summary} in~\hyperref[Appendix B]{Appendix B} summarizes the coding results. For our analysis we used the transcripts of group discussions and the written reports of all the 14 teams. See Table~\ref{tab:data_choice}. 

\subsubsection{DBT -- What the data reveals}\label{sec:V.B(ii)}

All the teams showed an awareness of design criteria and constraints relevant to their problem. Team-02, in their self driving golf cart design, stated their criteria and constraints explicitly:
\begin{quote}
~\hl{\textit{``To reiterate, our criteria is to compute the shortest possible travel time, and minimum work and net force required. This is to be done under the constraint of being safe''}}. 
\end{quote}
While coding, we considered whether the team intended to refer to safety as a criterion rather than a constraint. One perspective is that safety is a constraint because it represents a non-negotiable requirement that the design must meet. We had a productive discussion on this issue during coding and it is our view that educators take a flexible stance to accommodate students' styles of thinking and communication. In our experience, this is not an isolated instance of the blurry difference between what may be called a criterion or a constraint. Despite this challenge, these are useful notions to adopt in a design process.  

Only one team referenced economic aspects (criteria and constraints), unsurprising in a physics course but this might differ in an engineering context or if students were required to build a physical prototype. Likewise, fewer than half the teams mentioned safety aspects in their designs, indicating that these considerations were not a primary focus for most students.  For instance, Team-10 integrated metrics into their constraint, stating:
\begin{quote}
\hl{~\textit{``...bulletproof vests [need to] stop the bullet in 3.8 cm from a bulletproof vest, which is a constraint of the system''}}. 
\end{quote}

Almost all teams cited sources or referenced online materials (see Figure~\ref{fig:DBT_code_summary} in~\hyperref[Appendix B]{Appendix B}). A few teams mentioned gaining insights through discussions with their TA, while one team drew inspiration from classroom lectures. Additionally, some teams referenced national regulatory authorities, such as the~\textit{National Institute of Standards and Technology}~(Team-08) and the~\textit{National Highway Traffic Safety Administration} (Team-02). While it may be argued that referencing is typically associated with academic rigor rather than design thinking, the practice of referencing plays a complementary role in guiding students towards situating their work, justifying design choices, and informing iterative improvements. Although we did not find literature specifically addressing referencing within design thinking, we believe it serves as a valuable tool for refining ideas and linking them to established knowledge, thus promoting iterative design.

All teams incorporated numerical details relevant to their design. Team-12, whose problem was about roller-skates provided details within their problem statement: 
\begin{quote}
    \hl{\textit{``To align themselves with other roller skate companies like SkatePro, the mass of each Apollo Skate wheel is 500 g, their diameter is 60 mm
and their width is 34 mm.''}}
\end{quote}

\begin{figure}[H]
\centering
\caption{Team-12 provided numerical details related to their design directly in their python code.}
\fbox{\includegraphics[scale= 0.6]{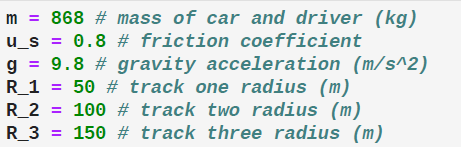}}
\label{fig:design_detail_team_12}
\end{figure}

Team-09 chose to include the numerical details while presenting their solutions, while Team-12 found it convenient to employ a python coding cell for this purpose (see Figure~\ref{fig:design_detail_team_12}).

\begin{figure}[H]
\centering
\caption{Team-09's Toy Catapult Design Mechanism. The illustration highlights both design elements and physics principles: the projectile is modeled as a point mass, while the launch mechanism is treated as an extended body.}
\fbox{\includegraphics[scale= 0.6]{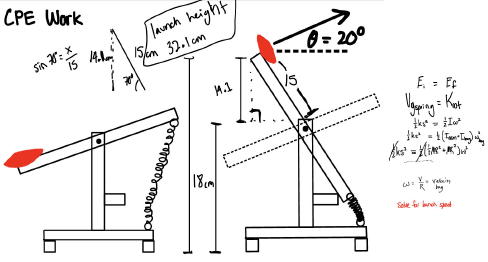}}
\label{fig:team_09_toy_catapult_mechanism}
\end{figure}

Similarly, in terms of design mechanisms, different teams demonstrated varying levels of detail as they progressed over the weeks. As mentioned in Section~\ref{sec:V.B(iii)}, some teams worked on contraption design, while some chose to detail processes. For instance, Team-09 represented their mechanism pictorially (Figure~\ref{fig:team_09_toy_catapult_mechanism}) while also incorporating numerical details of the dimensions of their contraption within their Python code.

In contrast, Team-07 worked on developing a process: 
\begin{quote}
\hl{\textit{``We are asked to develop a program that allows the delivery drones to travel destination with the
provided mail in-the most efficient and safe way.''}}
\end{quote}

In this study, students were required to develop an initial solution (first attempt) and iterate at least once, with additional iterations being optional. Our expectation was for students to engage in the process of generating solutions rather than focusing on achieving complete or perfect solutions. Given the time and resource constraints, this would not have been practical.

\begin{figure}[H]
\centering
\caption{Team-10's second iteration of their solution. The x-axis represents the flattened length, while the y-axis shows the force on the bullet. A higher force corresponds to a greater reduction in length.}
\fbox{\includegraphics[scale= 0.4]{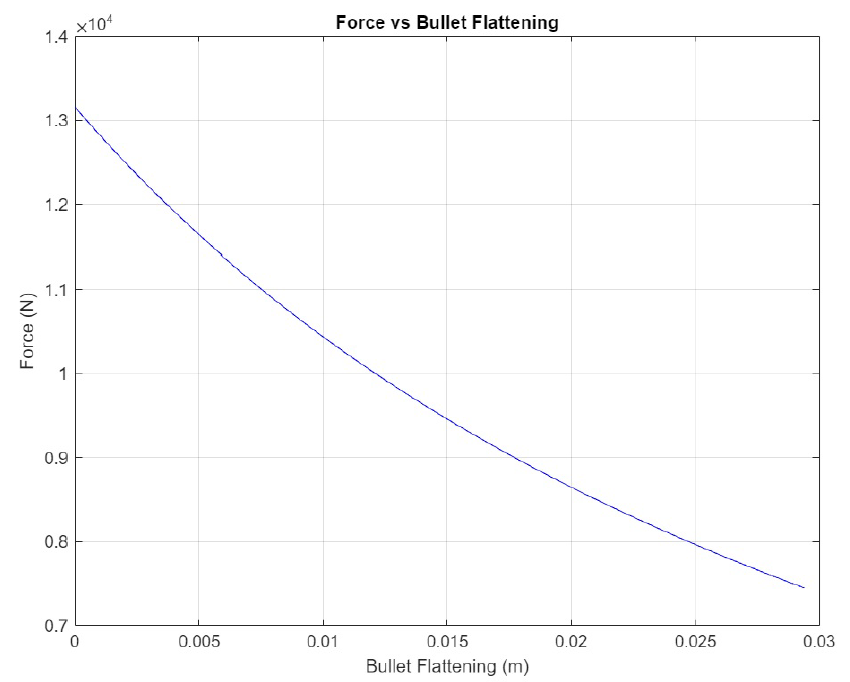}}
\label{fig:team_10_graph}
\end{figure}

Most teams presented detailed solutions in their first attempt and refined them in the first iteration. While some teams made only minor modifications, others engaged in substantive refinements, incorporating additional calculations, justifications, or alternative design choices. For example, Team-10 initially manually calculated the force exerted by a bulletproof vest on an incoming bullet, assuming a set of parameters. In their first iteration, they incorporated bullet flattening and automated calculations using Python. By their final iteration, they generated a graph (Figure~\ref{fig:team_10_graph}) depicting the force on the bullet as a function of its flattening due to impact, also highlighting the level of comfort in using Python coding to automate calculations. 

However, not all teams remained fully engaged in the task. Team-11 made only cursory entries in the first attempt and second iteration, preferring to perform most calculations manually rather than using Python (see Figure~\ref{fig:python_inductive_coding} in~\hyperref[Appendix C]{Appendix C} and Figure~\ref{fig:DBT_code_summary} in~\hyperref[Appendix B]{Appendix B}). A similar pattern was observed in Team-07.

In their reports, students were asked to comment on the limitations of their solutions after each iteration. However, only about half of the teams responded, and those who did provided only general statements. A smaller fraction of teams discussed limitations during their conversations, though with even less detail. Team-10 identified not having considered the effect of bullet flattening as a limitation in their initial attempt. They analyzed a scenario where the bullet flattened to half its original length but acknowledged that this assumption was not representative of real-world cases, which they recognized as a limitation. Similarly, Team-11 assumed that bungee cords obey Hooke’s Law, later noting that this may not truly represent the behavior of cords in practice, which they recognized as a limitation in their final solution. These responses suggest that students needed more structured guidance on how to critically assess and articulate the limitations of their design solutions.

We observed variations in how teams approached iterations. While some engaged in cursory or token iterations, a smaller fraction pursued meaningful refinements. Given that Week 14 was the final week of the semester, some teams may have experienced end-of-semester fatigue~\cite{fatigue_law_2007}. Moreover, more iterations did not always lead to better solutions, highlighting the need for a more systematic approach to guiding students in the iteration process.

In our rubric (see~\hyperref[Appendix D]{Appendix D}), we introduced the notion of trade-offs, a key consideration in tackling design problems. However, only a few teams explicitly addressed this aspect. It appears that we were not entirely successful in conveying its intended meaning. Some responses were vague, such as Team-11’s statement:
\begin{quote}
\hl{\textit{``trade-off we used was energy principle for momentum principle.''}}
\end{quote}

In contrast, Team-12 articulation was slightly better:
\begin{quote}
\hl{\textit{``one trade-off we wanted to highlight is the fact that first we wanted to try and solve for the velocities using all four wheels but we realized that the velocity of one wheel would be sufficient as well as the fact that solving for all the four wheels would over complicate the problem.''}}
\end{quote}

Team-09, kept it shorter and was to the point: 
\begin{quote}
\hl{\textit{``The light-weight material of the toy (mostly plastics) can cause the toy to be fragile
unless made from more expensive types of plastic. The safety features will also partially inhibit the movement of the catapult.''}}
\end{quote}

Ensuring all teams make meaningful progress in activities such as this remains a common challenge in large classes, where lab sessions are often managed by a single TA, occasionally with the assistance of an undergraduate teaching assistant. In a section of about 50 students, as in this study, this remains a significant challenge in interacting with every group and offering detailed suggestions. Although many TAs provide feedback while grading, this is equally challenging given the number of sections they teach.

Overall, though there is much scope for improvement, students' approach to design thinking best aligns with Crilly, who defines design thinking as:~\textit{``a disciplinary approach adopted by designers, including product and service designers. These are people who devise plans to bring new things into existence or to improve situations (often with a focus on human experience)''}~\cite{crilly_design_2024}.


\subsubsection{DBT -- Moving forward}\label{sec:V.B(iii)}

To enhance students' design-based thinking in future iterations, we propose a more detailed rubric, though we acknowledge the risk of it becoming a checklist for students to simply tick off.

Additional guidance on the meaning and application of terms like criteria, constraints, and trade-offs within the context of their designs could help deepen their understanding. Encouraging students to reflect on how and why they iterated may foster a more thoughtful design process; a reflective evaluation could be valuable here. Asking students to engage in role play, where one person acts as the client, could facilitate debate and encourage more critical examination of their solutions.

In our observations, some teams made only minor adjustments in their iterations, without critically assessing their solutions. A stronger emphasis on iterative refinement, trade-off analysis, and consideration of real-world constraints would help them adopt a more rigorous approach. Incorporating hands-on prototyping, alongside encouraging the use of freely available databases for data analysis or running simulations, could further enrich their iterative design process. 

In our recent publication, we emphasized the importance of asking students to engage in a~\textit{feasibility study}~\cite{ravi_prper_2024, feasibility_thabane_tutorial}. We realize that this term was never introduced during our instructions, and it is certainly an area where we can improve.

\subsection{Science-Based Thinking (SBT)}\label{sec:V.C}

In this section, we examine how science-based thinking (SBT) is reflected in students' work. Section~\ref{sec:V.C(i)} outlines our method for analyzing the data, Section~\ref{sec:V.C(ii)} highlights key insights, and Section~\ref{sec:V.C(iii)} offers strategies to support the development of students' science-based thinking.

\subsubsection{SBT -- Examining the data}\label{sec:V.C(i)}
In our rubric (see~\hyperref[Appendix D]{Appendix D}), we asked students to engage in what we referred to as \textit{science thinking}~\cite{ravi_prper_2024}, which included the following components: (i) the use of at least two of the three fundamental principles—momentum, energy, and angular momentum—along with a justification for their relevance; (ii) identifying the assumptions and approximations made; (iv) identifying the system and surroundings specific to the problem (systems thinking); and (v) employing modeling by explicitly stating and justifying the use of a `model' (e.g., point-particle) in the solution. Since teams did not detail upon the choice of their model directly, for the purposes of this study, we found it suitable and sufficient to merge point (v) into (iv) under `systems and surroundings'. Furthermore, points (iv) and (v) involve deeper notions such as `systems thinking'~\cite{monat2015systems, salado2019systems, cabrera2023systems} and `modeling' in science~\cite{brewe2008modeling, hestenes1992modeling}, the discussion of which is well beyond our immediate pedagogical goals.

\begin{figure}[H]
\caption{One of the iterations of the coding sheet for analyzing students' science-based thinking is presented below. The final results may be seen in~\hyperref[Appendix B]{Appendix B}.}
\fbox{\includegraphics[width=0.96\linewidth]{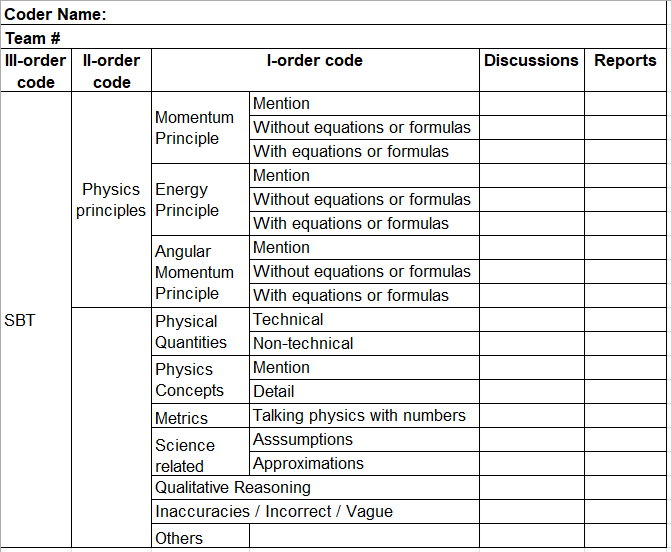}}
\label{fig:code_sheet_sbt_blank}
\end{figure}

An iteration of the coding sheet used by our research team may be seen in Figure~\ref{fig:code_sheet_sbt_blank}. The coders would take notes in the sheets independent of each other, discuss, and arrive at a consensus. The process was repeated for each of the 14 teams for both transcripts of group discussions and the written reports (see Table~\ref{tab:data_choice}).  The final consolidated coding results displayed in Figure~\ref{fig:SBT_code_summary} in~\hyperref[Appendix B]{Appendix B}. 

In this study, we distinguish between physics principles and physics concepts, although some overlap is inevitable. By~\textit{physics principles}, we specifically refer to the three fundamental principles: momentum, energy, and angular momentum that are presented in the textbook~\cite{chabay2015matter} used in this course. In contrast,~\textit{physics concepts} include terms such as force, air resistance, and buoyancy, among others. We find it useful to adopt this distinction as our course textbook~\cite{chabay2015matter} follows a principle-based approach. Some might view this as a bookkeeping measure. Although our approach was informed by our recent study~\cite{ravi_prper_2024}, the reader will notice a significant increase in the level of detail in the coding schema for this study.

\subsubsection{SBT -- What the data reveals}\label{sec:V.C(ii)}

While only a couple of teams addressed the notion of `system and surroundings' within their problem statements, a majority of teams addressed this while progressing through their solutions. Team-12, in the course of writing down the related equations, explicitly stated:
\begin{quote}
\hl{\textit{``We started by using an extended system with the person and the skates in the system and the hill and Earth as the surroundings.''}}
\end{quote}
This amount of detail is crucial, as it helps us examine how they applied physics concepts and principles to their system. However, a robust understanding of a system also necessitates consideration of its surroundings—an aspect that only a smaller fraction of students explicitly addressed in their solutions.

Group discussions across teams tended to be more general and, at times, reiterated points that were documented in the reports. However, the value of these discussions should not be overlooked, as they provide insight into how ideas evolved within teams and highlight the interests and contributions of different members -- an aspect which may be of interest for some researchers. In contrast, the written reports, while lacking the dynamic nature of discussions, exhibited a higher level of detail, reflecting a more structured engagement with the design process. For example, Team-01, during group discussions, reflected on their iterations:
\begin{quote}
~\hl{\textit{``And then from there we iterated through multiple degrees for the first part and then a multitude of impact locations for the second part and graphed our findings''}}, further adding~\hl{\textit{``the confidence in our analysis spreads to two primary aspects of baseball. The optimum angle and the point of impact on the ball. By understanding these factors, we’re going to improve the performance and make more informed decisions and adjust our techniques and yield better results''}}. 
\end{quote}
Evidently, the team is trying to convey that the range of the projectile depends on both the angle of projection relative to the horizontal and the point where the bat impacts the ball, and they outline how they intend to calculate the projectile's range. Although the conversation snippet may seem a bit unclear, the team presented calculations to support their reasoning in their reports by employing the momentum update principle employing Python code (for the code, see Figure~\ref{fig:CT_stud_sample_03}). 


We expected students to apply at least two of the three fundamental principles but allowed flexibility to accommodate their natural approaches to applying physics concepts. A majority of teams applied the momentum principle and/or the energy principle, while fewer teams incorporated the angular momentum principle (see Figure~\ref{fig:SBT_code_summary} in~\hyperref[Appendix B]{Appendix B}). This may be attributed to the nature of the problems chosen by teams or the fact that angular momentum is introduced later in the course (around week 10), leaving students with less time to assimilate and apply related ideas. Team-07 outlined their use of momentum principle incorporating the use of mathematics as follows:
\begin{quote}
\hl{~\textit{``The equations and expressions we’ll be using are the momentum principle which is the net force multiplied by the change in time. We will measure the force on the drone over the given time while varying the mass being carried by the drone. The mass of the drone will always remain constant for this we’re using the constant variable of 35 kg and once we determine the force we can plug this into our equation of momentum to solve for the momentum of the drone over the given period of time which is the time it takes to deliver the package''}}. 
\end{quote}
In progression toward their solution, in their reports, they generated a graph and table adding a level of detail to their discussions. 

Team-11 explicitly articulated their assumption (negligible air resistance) and approximation (the cord obeys Hooke’s law), along with their initial conditions:
\begin{quote}
\hl{~\textit{``We assume air resistance is negligible and the cord obeys Hooke’s law, and that the person will start and end at rest''}}. They further clarified their system and surroundings, stating:\hl{~\textit{``The system is the person, cord, and the Earth, and we have nothing in the surroundings because no work is done on the system''}}. 
\end{quote}
This statement implicitly invokes the principle of mechanical energy conservation, demonstrating their recognition of how system boundaries influence energy analysis. Similarly, Team-14 outlined their application of the angular momentum principle to derive an expression for the angular speed of the yo-yo as it descends: 
\begin{quote}
\hl{~\textit{``We are going to use the relation that torque = I [moment of inertia] times alpha, and torque equals r cross F. So we’re going to find the torque that the string exerts on the yo-yo, which is our force times its radius. And then we’re going to use the fact that the angular acceleration equal torque over moment of inertia. That will give us the angular acceleration and the angular velocity of the yo-yo, which tells us how fast the yo-yo is moving and how fast it is speeding up, which is what the toy companies wanted''}}. 
\end{quote}
Notably, the reference to stakeholders (toy companies) highlights students' awareness of the design context, making evident how students draw upon various ways of thinking quite unpredictably. 

Qualitative reasoning was frequently observed, though some teams made inaccurate statements or misinterpreted results, indicating a need for more careful articulation. Team-01 explained the absence of torque on a (uniform) ball when struck by the baseball bat such that the line of action of the force passes right through the geometric center of the mass as:
\begin{quote}
~\hl{\textit{``When it comes to where to hit the ball, if we strike dead center of the ball (r perpendicular would be 0), we have no angular momentum, and so everything will be purely translational''}}. 
\end{quote}
Similarly, Team-06 presented an interesting observation about airflow patterns in Formula 1 racing, explaining how they create a downward force equivalent to increasing the vehicle’s load and traction: 
\begin{quote}
\hl{\textit{``Grip is similar to friction, and is dependent on the mass of the car and the amount of down-force that the car can generate, (which artificially adds mass to the car at certain speeds). Down-force is generated through the use of aerodynamic surfaces that funnel air into the sky, thereby pushing the car into the ground.''}}.
\end{quote}

One notable concern was the lack of attention to numerical precision. None of the teams presented numerical results with appropriate significant figures, likely due to an over-reliance on Python for automated calculations. Additionally, since students were not creating physical models or working with data from direct experimentation, they may not have perceived significant figures as particularly relevant. This combination of factors likely contributed to the oversight in numerical precision. Additionally, a significant number of teams neglected to include correct units when reporting results—an essential expectation in scientific communication. The numerical results presented by Team-02 (see Figure~\ref{fig:CT_stud_sample_01}) exemplify this trend, which was prevalent across almost all project reports (see Figure~\ref{fig:SBT_code_summary} in~\hyperref[Appendix B]{Appendix B}).

Given that this is a physics course, we paid particular attention to instances where students might have applied concepts incorrectly, held misconceptions, or made inaccurate statements. At times, we wondered if it was simply oversight. We identified several notable examples that highlight these areas. Team-01, while generating a Python code for momentum updates (see Figure~\ref{fig:CT_stud_sample_02}), incorrectly commented within their code that\hl{~\textit{``momentum is conserved''}}. Given that their python code is correct, we interpreted this to be likely an oversight. Moreover, they did not clearly define their system, which leaves room for ambiguity and potentially misleading conclusions. Team-03, while modeling the mortar shell as a cylinder, stated:
\begin{quote}
\hl{~\textit{``...we found what the moment of inertia was that would be 1/12 M L squared because we were approximating that the mortar was almost a cylinder''}}. 
\end{quote}
While they calculated the moment of inertia, they did not specify the axis around which the calculation was made, which is an important detail for accuracy in physical modeling. It is possible it was another instance of oversight. Similarly, Team-14, when analyzing the forces on the yo-yo, stated:
\begin{quote}
\hl{~\textit{``Then in order to find the acceleration of the yoyo at the bottom, we’re going to use the fact that the yoyo is now still, so the force of tension up will equal mg, the weight of the yoyo down, when it will stay in place''}}. 
\end{quote}
This suggests the~\textit{conceptual resources}~\cite{hammer2000student} that students are activating to describe the forces at play at the lowest point of the yo-yo's motion. In reality, the net force is not zero at the lowermost point. It appears the students may be relating momentary rest as implying zero acceleration. For a more accurate understanding of the yo-yo's behavior, we refer the reader to the insightful article by Minkin and Sikes, which explains the phenomenon of how a yo-yo~\textit{sleeps} at the lowest point of its trajectory~\cite{yoyo_minkin_2020}. This example underscores the complexity of real-world problems, demonstrating that even seemingly straightforward scenarios can invite deeper investigation and further research~\cite{yoyo_izara_2011}. Despite this conceptual oversight, Team-14 did demonstrate a correct understanding in another aspect, stating that\hl{~\textit{``the yo-yo won’t slip on the string''}}, which is a key insight to use the rolling constraint equation $v_{\text{cm}} = r \omega$. 

In our previous study~\cite{ravi_prper_2024}, we advocated for strengthening the \textit{design-science connection}. A strong understanding of science concepts can enhance design, while weak scientific reasoning can derail it. Engaging in a critical evaluation of results may help mitigate such issues.
For instance, Team-05 determined a spring constant (about 3500 N/m) which they thought was unrealistically high stating: 
\begin{quote}
\hl{\textit{``roughly 3 - 4 times more than a heavy bow shot would be pulled''}},
\end{quote}
but demonstrated thoughtfulness by referring to online sources and acknowledging the discrepancy. In contrast, Team-09, despite having a well-conceived design, used incorrect formulas, leading to misleading results that they did not verify. They also assumed a spring constant of 100 N/m, which seemed excessively high for a toy. Similarly, Team-12, though correctly applying physics concepts, made a calculation error that resulted in at least two issues: (1) a formula that was dimensionally incorrect, which they failed to notice, and (2) a numerical result for the skater's speed (about 11 m/s) that exceeded that of an Olympic sprinter! Not to be left behind (pun intended), Team-02 calculated their golf cart’s speed to be around 9 m/s but did not critically evaluate the plausibility of such a high speed. To be fair to the students, it is possible that they did not notice these issues due to lack of sufficient time to review their calculations and reflect on their results.

The importance of deep understanding of physics concepts is clear in these examples. Without careful attention to accuracy and clarity, even well-intentioned design efforts can falter, making it essential for students to avoid conceptual flaws and ensure their scientific understanding underpins their design work.

\subsubsection{SBT -- Moving Forward}\label{sec:V.C(iii)}

We can support students by encouraging validation of their results through peer discussions or instructor feedback. This feedback loop could help identify and correct inaccuracies in reasoning and calculations, fostering a more rigorous approach to problem-solving. 

Framing of ill-structured problems is not easy, and solving is even more so. Considering this, students may be encouraged to engage in `systems thinking'~\cite{ravi_SWoT_01_2024} -- to look at the problem as an ensemble of multiply connected parts and then consider solving that aspect which may yield to the application of physics concepts and principles. This may be a more pragmatic approach given the limited time resources within laboratory settings. 

Another intervention would be to place a stronger emphasis on numerical precision, particularly regarding significant figures and units. Integrating error analysis into weekly hands-on lab activities would enhance students' awareness of precision and accuracy in physics-based problem-solving. Explicitly requiring students to focus on these aspects in their reports would reinforce best practices in scientific communication. 

Given that this is a physics course, a systematic approach to iterations could include requiring students to explain whether each iteration meant an incremental shift in the application of physics concepts, thereby leading to a better solution.  

To deepen their understanding of how theoretical concepts apply to real-world phenomena, we can guide students to engage in data-driven analysis. Given the constraints on resources that limit hands-on or physical model creation, incorporating openly accessible real-world datasets from sources such as NASA's~\textit{Earth Observing System Data and Information System}~\cite{nasa_earthdata} (EOSDIS), the~\textit{U.S. Geological Survey}~\cite{usgs_lpdaac} (USGS), or the~\textit{National Oceanic and Atmospheric Administration}~\cite{noaa_opendata} (NOAA), students would gain practical experience and develop critical thinking skills, connecting physics principles to actual data, all while overcoming the limitations posed by resource availability.

\subsection{Mathematics-Based Thinking (MBT)}\label{sec:V.D}

In this section, we investigate how mathematical thinking is demonstrated in students' work. Section~\ref{sec:V.D(i)} describes our approach to analyzing the data, Section~\ref{sec:V.D(ii)} presents key observations, and Section~\ref{sec:V.D(iii)} suggests strategies to foster students' mathematical reasoning.

\subsubsection{MBT -- Examining the data}\label{sec:V.D(i)}

Our rubric (see~\hyperref[Appendix D]{Appendix D}) prompted students to engage in what we referred to as~\textit{mathematical thinking}~\cite{ravi_prper_2024}, emphasizing (i) supporting findings or results with numerical calculations and (ii) utilizing graphs, equations, and tables. This guidance was informed by the lead author's preliminary investigations~\cite{subramaniam2023narst, ravi_perc_2023}.

\begin{figure}[htbp]
\caption{Code Summary for Mathematics-Based Thinking (MBT)}
\fbox{\includegraphics[width=0.9\linewidth]{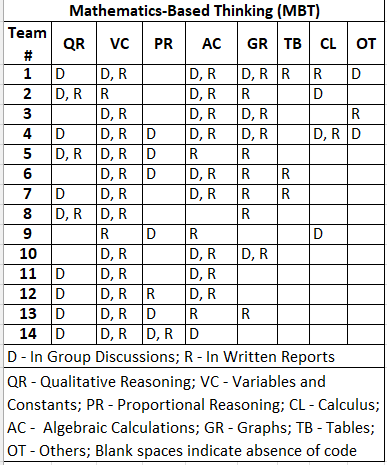}}
\label{fig:MBT_code_summary}
\end{figure}

For our analysis we used the transcripts of group discussions and the written reports of all the 14 teams. See Table~\ref{tab:data_choice}. The MBT coding sheet was similar to what may be seen in Figure~\ref{fig:code_sheet_sbt_blank} and the coding results are summarized in Figure~\ref{fig:MBT_code_summary}. The coding scheme, drawing from our recent study~\cite{ravi_prper_2024} and the first paper~\cite{ravi_SWoT_01_2024} in this series, was developed to align with our pedagogical goals. Specifically, we examined students' use of mathematical equations, their derivation of formulas grounded in physics principles, and their integration of both qualitative and quantitative reasoning. These elements provide insights that will inform future instructional interventions. 

Since teams used Python for calculations, readers may notice an overlap with computational thinking, which we address separately in Section~\ref{sec:V.F}. However, this overlap will not hinder our discussion, as our focus here remains on mathematical thinking.

\subsubsection{MBT -- What the data reveals}\label{sec:V.D(ii)}
For MBT, we analyzed students' use of algebra, including basic trigonometry and geometry, in their calculations. Despite this being a calculus-based course, only one team (Team-04) explicitly applied calculus in their analysis. This team formulated coupled first-order differential equations in two dimensions to account for air friction in projectile motion, implemented a Python script to solve the equations, and graphically represented the results (see Figure~\ref{fig:CT_stud_sample_04}). Notably, this teams' Python implementation and mathematical approach extended well beyond the curricular expectations, illustrating how students can sometimes exceed anticipated learning boundaries. 
A smaller fraction of teams employed an iterative approach to updating position and momentum. While the calculations may initially appear algebraic, they implicitly engage with principles of differential calculus (see Figure~\ref{fig:CT_stud_sample_03}).

\begin{figure}[htbp]
\caption{Following qualitative reasoning with detailed mathematics. An example from Team-04.}
\fbox{\includegraphics[width=0.9\linewidth]{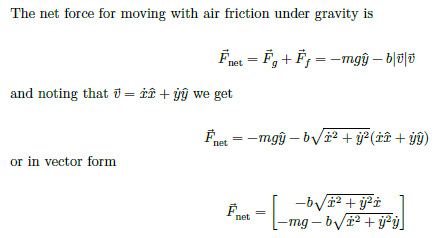}}
\label{fig:team_04_diff_eqn}
\end{figure}

While there were instances of qualitative reasoning during discussions in about half of the teams, only a smaller fraction followed up with additional, detailed calculations in their reports. Team-04 articulated their reasoning behind using differential equations as follows:
\begin{quote}
\hl{~\textit{``I think the air resistance was like...just complicated you know…it’s based on total velocity and not just directional velocity and that’s really hard especially when you’re considering that gravity acts with it on way up and against it on way down and it varies constantly and if you try and write it out normally you almost see it as a recursive equation because it’s like calculating the velocity requires you to use a function that is based on the velocity and it almost seems to be near impossible without either simplifying it or by using math tricks.''}} 
\end{quote}
This team extended their qualitative reasoning into detailed mathematical formulations (see Figure~\ref{fig:team_04_diff_eqn}) and generated graphs using Python code in their reports (see Figure~\ref{fig:CT_stud_sample_04}).

This example highlights the importance of capturing multiple forms of student data—while group discussions have the potential to reveal teams' conceptual struggles and reasoning processes, written reports often showcase a more structured and mathematically rigorous articulation of their ideas. Additionally, beyond MBT, this conversation snippet illustrates how other elements of our~\textit{WoT4EDP} framework emerge naturally. Students draw on their physics knowledge of gravity and drag force (SBT) while also reflecting on the challenges inherent in the problem (MER). This reinforces the idea that students’ thinking is far too complex to be neatly categorized into isolated domains, emphasizing the interconnected nature of their reasoning. 

We also observed a significant overlap between MBT and computational thinking (CT)~\cite[p.146]{shute2017demystifying}, as students leveraged Python to generate graphs and automate calculations that would have been tedious manually (see Figures~\ref{fig:CT_stud_sample_03} and~\ref{fig:CT_stud_sample_04}). While most teams used Python for graph generation, only a small number utilized it to create tables. Nearly all teams referenced variables and constants in their Python code (see Figures~\ref{fig:CT_stud_sample_01} and~\ref{fig:CT_stud_sample_02}). Only a small fraction (see Figure~\ref{fig:MBT_code_summary} of teams demonstrated proportional reasoning, with more instances appearing in discussions than in reports (see Figures~\ref{fig:CT_stud_sample_01} and~\ref{fig:CT_stud_sample_02}). For instance, Team-14 articulated how a lighter Yo-Yo would achieve a higher speed due to its lower moment of inertia, stating:
\begin{quote}
\hl{~\textit{''...it does prove our point that the lower mass Yo-Yo does have a lower [moment of] inertia [about an axis through its center and perpendicular to its plane] which results in a faster spin''}}. 
\end{quote}
The reader may notice how students' language weaves together both physics and mathematics. Notably, this team did not explicitly specify the axis when referring to the moment of inertia, either in their discussions or reports. This example further reinforces our choice of data sources —group discussions and reports. Our experience analyzing group discussions supports the value of taking a more dialogic approach to promote student learning, ultimately fostering a~\textit{``culture of talk''}~\cite{firetto2023embracing} in our classrooms. Just as writing supports~\textit{Writing to Learn} (WTL)~\cite{hoehn2020framework}, our findings suggest that discussions have the potential promote what we call~\textit{Talking to Learn} (TTL). Further research is needed to develop scaffolds and frameworks that foster Hammer's notion of \textit{``productive''}~\cite{hammer1995student} discussions,  or \textit{``rich talk''} as put forth by City~\cite[p.13]{city2014talking}. 

\subsubsection{MBT -- Moving Forward}\label{sec:V.D(iii)}
A key gap in students’ MBT was the limited application of calculus in their problem-solving, despite its relevance to the course. While some teams implicitly used differential calculus concepts through iterative position and momentum updates, there was little explicit engagement with calculus-based reasoning. Python was primarily used for graphing, with very few teams employing it to generate tables.

To address these gaps, targeted instructional strategies can be implemented in future iterations. Inclusion of calculus-based tasks in the course, or in laboratory tasks if possible, could help students recognize its relevance and applicability. Incorporating assignments that require students to use Python for both graphing and tabular data representation could encourage broader engagement with computational tools. Additionally, students should be encouraged to analyze graphs and tables more deeply, identifying patterns and discussing the underlying science to better inform their design decisions.or patterns, and discussing the underlying science so as to inform their design decisions.  

To strengthen proportional reasoning, guided reflection activities or structured peer discussions could help students articulate and apply mathematical relationships more effectively. Finally, embedding opportunities for students to collect, analyze, and integrate experimental or database-sourced data into their projects could foster deeper engagement with real-world problem-solving and reinforce the role of empirical data in mathematical modeling.

In future iterations of our ED-based project, to further support MBT, we see scope for integrating elements of the~\textit{ACER} framework proposed by Wilcox {\em et al.}~\cite[p.4]{wilcox2013analytic}:~\textit{Activation}—problem cues trigger relevant mathematical tools;~\textit{Construction}—students develop simplified mathematical models using equations and representations;~\textit{Execution}—students manipulate mathematical expressions while considering their physical meaning;~\textit{Reflection}—students evaluate their results, checking for errors and assessing physical insights. 

\subsection{Metacognitive Reflection (MER)}\label{sec:V.E}

In this section, we explore how students engaged in metacognitive reflection. Section~\ref{sec:V.E(i)} explains our selection of data and the method used for analysis, Section~\ref{sec:V.E(ii)} highlights key insights, and Section~\ref{sec:V.E(iii)} suggests strategies to foster productive reflections that enhance students' problem-solving skills.

\subsubsection{MER -- Examining the data}\label{sec:V.E(i)}

While reflective statements appeared in both transcripts (see Figure~\ref{fig:wot_disc_chart_1}) and reports, our focus on future pedagogical iterations led us to identify the most relevant portion of data for analysis. In the conclusion section of their final reports, we asked students:

\begin{quote}
\textit{Write about your final solution to the problem. Were you able to fully address all aspects of the problem as initially planned?}
\end{quote}

Student responses to this prompt provided valuable insights, as written reports represented collective reasoning of the team rather than individual contributions in discussions. Our goal was to assess whether this prompt elicited meaningful reflections that could inform future course design. Hence, we used teams' responses to the above prompt from the written reports of all 14 teams. See Table~\ref{tab:data_choice}. 

\begin{figure}[htbp]
\caption{Code summary for Metacognitive Reflection (MER). Blank spaces indicate `No' while `Repeat' denotes that teams merely restated what they had previously mentioned in the report.}
\fbox{\includegraphics[width=0.9\linewidth]{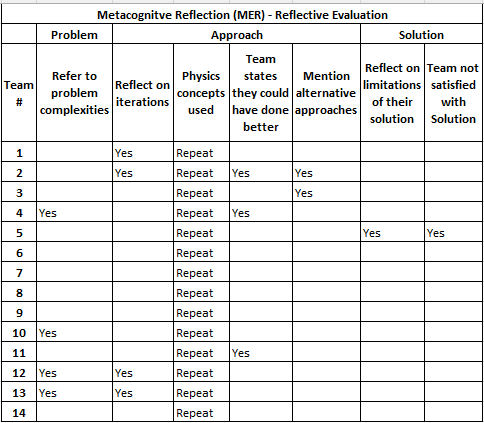}}
\label{fig:MER_code_summary}
\end{figure}

Our decision was also influenced by Stanley and Lewandowski's recommendations for~\textit{authentic scientific documentation}~\cite{stanley2018recommendations}, which align with our interest in how students justify design decisions in written form. Rather than applying a predefined framework, we allowed insights to emerge directly from the data. See Figure~\ref{fig:MER_code_summary} for the coding sheet.

\subsubsection{MER -- What the data reveals}\label{sec:V.E(ii)}
Most teams either restated their results or summarized their solution approach.  All teams merely repeated what they had stated earlier in the reports, the variation being only in the length and detail of their responses, with nothing new to be gleaned. 

Only a very small fraction of teams mentioned their approach could have been better. Only one team appeared not satisfied with their solution. Just one team explicitly mentioned limitations of their solution.

Only a smaller fraction of teams (see Figure~\ref{fig:MER_code_summary}) recognized the complexities of real-world problems, adding why that prompted them to make adjustments along the way. Team-12, which initially aimed to determine the final speed of a person on roller skates moving downhill, explained why they had to scale down their goals after realizing their original problem was too complex. They reflected:
\begin{quote}
\hl{~\textit{``Our problem altered a lot throughout the process. For instance, we initially wanted to calculate the velocity of all four wheels combined but realized that real-world physics problems are actually quite difficult to solve. Therefore, we decided to make our situation more manageable by solving for the velocity of just one of the four wheels instead''}}. 
\end{quote}
This reflection is particularly valuable as it highlights a common challenge students face—setting ambitious goals initially but needing to scale back due to time constraints and the complexity of real-world problems. Similarly, Team-10 reflected on the limitations of their design for a bulletproof vest, recognizing that despite their calculations, the final product might be\hl{~\textit{``too bulky and unusable''}}. This realization underscores how focusing solely on physics was insufficient to address all aspects of the problem, and they demonstrated awareness of this limitation. Team-04 also acknowledged the need to simplify their approach:
\begin{quote}
\hl{~\textit{``Our problem involved too many moving parts for us to solve within the scope of our Phys 172 curriculum. We pivoted and simplified it into a problem we were equipped to handle''}}. 
\end{quote}
Meanwhile, Team-02 and team-03 wondered how their results might have changed had they made different assumptions, indicating some level of reflection on the problem-solving process.

In summary, students' responses to this question provided little new insight into the teams' thinking, beyond what had already been stated by the teams. It is possible that they were fatigued by the end of the report~\cite{fatigue_2023} or that the wording of the prompt did not effectively encourage deeper reflection.

\subsubsection{MER -- Moving forward}\label{sec:V.E(iii)}
A more structured approach to fostering metacognitive reflection may be necessary, as simply asking students to reflect on their solutions does not appear to be effective. As Hacker puts it, students~\textit{``should be directed toward specific, higher-level analyses''}~\cite[p.79]{hacker2003not}. MER has the potential to reveal the interconnectedness of students' Ways of Thinking. While interconnectedness of the ways of thinking is not intentionally explored in this paper, it will be the focus of our third paper. For the next iteration, we plan to use prompts specifically designed to elicit this interconnectedness, encouraging students to critically assess their solutions by considering physics, math, coding, assumptions, feasibility, and innovation. This aligns with our goal of fostering deeper metacognitive engagement and justification of design decisions. We also see scope for incorporating ideas from~\textit{Science Writing Heuristic} (SWH)~\cite[p.5]{hoehn2020framework}, which encourages reflective writing at the end of a lab report.

\subsection{Computational Thinking (CT)}\label{sec:V.F}

In this section, we explore how students engaged in Computational Thinking (CT). Section~\ref{sec:V.F(i)} explains our selection of data and the method used for analysis, in Section~\ref{sec:V.F(ii)}, we present our~\textit{inductive} approach and findings, while in Section~\ref{sec:V.F(iii)}, we describe our~\textit{deductive} approach and results. Finally, in Section~\ref{sec:V.F(iv)}, we summarize our findings and provide instructional recommendations to enhance computational thinking.

\subsubsection{CT -- Examining the Data}\label{sec:V.F(i)}

In this study, we focus primarily on students' use of Python code, though it is important to recognize that Computational Thinking (CT) encompasses a broader scope, extending beyond these specific programming tools~\cite{ravi_SWoT_01_2024}. While our approach (CT-lite, in some sense) is narrowed, it aligns with our instructional objectives, particularly in fostering the development of students' CT skills in future iterations of our course.

There were only a few references to Python code in the group discussions (see Figure~\ref{fig:wot_disc_chart_1}), so we focused solely on the written reports of all 14 teams, as these included the Python code (see Table~\ref{tab:data_choice}). To examine how CT manifested in students' work, we employed a dual coding approach. 

First, we used an~\textit{inductive} coding approach to analyze the data from the ground up, allowing themes and patterns to emerge naturally from the students’ (python) coding practices and project outputs. This inductive method provided insight into the spontaneous CT skills students utilized, offering a baseline understanding of how our instructional design impacted their computational engagement. A coding sheet, similar to that in Figure~\ref{fig:code_sheet_sbt_blank}, was used in the early stages but is not reproduced here. The final consolidated coding sheet may be seen in Figure~\ref{fig:python_inductive_coding} in~\hyperref[Appendix C]{Appendix C}. 



Second, we applied a~\textit{deductive} coding approach, guided by the CT framework outlined by Shute {\em et al.}~\cite{shute2017demystifying}, which provided a structured lens through which to interpret specific elements of CT. This deductive layer enabled a systematic assessment of how closely students' work aligned with established CT principles, facilitating comparison with standard CT benchmarks. A coding sheet, similar to that in Figure~\ref{fig:code_sheet_sbt_blank}, was used in the early stages but is not reproduced here. Table~\ref{tab:deductive_CT_coding} in~\hyperref[Appendix C]{Appendix C} produces a summary of findings for the entire data set. 

By combining these approaches, we aimed to capture both emergent patterns in students' computational practices and assess the influence of our pedagogical interventions. Insights from this analysis will inform future iterations of our teaching strategies, with a focus on enhancing students' abilities to engage in complex computational tasks, deepen their understanding of CT within physics contexts, and enhance their design solutions.

\subsubsection{CT -- Inductive Coding}\label{sec:V.F(ii)}
The students’ engagement with CT across different engineering design-based physics projects displayed a range of approaches, skill levels, and degrees of reliance on automation. Although students were encouraged to use Python, we did not mandate it.

While standard Python libraries, such as~\lstinline{numpy} (used for efficient numerical calculations of vectors and matrices),~\lstinline{pandas} (used for data manipulation and analysis), and~\lstinline{matplotlib} (used for data visualization in graphs and charts), had been introduced in prior labs, only a few teams utilized~\lstinline{numpy} and~\lstinline{pandas}, whereas~\lstinline{matplotlib} (see Figure~\ref{fig:CT_stud_sample_04}) saw a relatively higher adoption rate. Notably, only one team applied an advanced library like~\lstinline{scipy} (used for advanced computations involving differentiation and integration), suggesting that deeper CT practices were less common. These patterns indicate that, while students demonstrated a baseline familiarity with Python's tools for data handling and visualization, advanced computational approaches were relatively rare.

Projects also demonstrated varying levels of computational sophistication. Some teams (see Figure~\ref{fig:CT_stud_sample_03}) displayed advanced coding practices, using functions (reusable custom code blocks defined by the~\lstinline{def} statement), loop structures (used for iterative execution of the same piece of code such as~\lstinline{while} and~\lstinline{for}), and conditional statements (the so-called~\lstinline{if-then} statements, used for controlled decision making in computer code) despite not having been formally taught these constructs in the laboratory course. These practices indicate that some students were either drawing on prior knowledge or independently exploring Python’s capabilities, revealing a diversity of computational thinking.

\begin{figure}[htbp]
\caption{A code snippet from Team-02 showing how students used Python for simple calculation tasks. The reader may also note how the numerical output has no units and runs to several digits, a trend observed in almost all teams.}
\fbox{\includegraphics[width=0.96\linewidth]{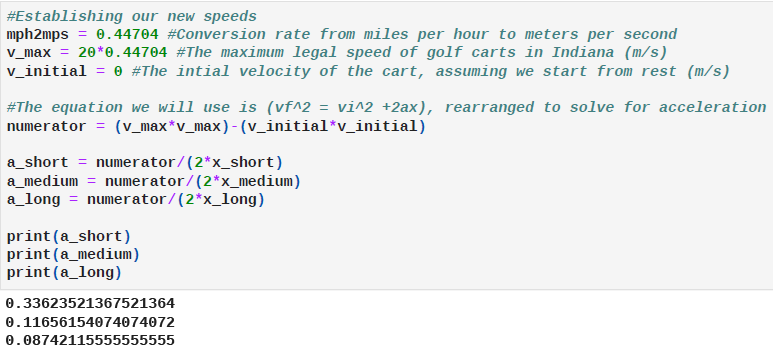}}
\label{fig:CT_stud_sample_01}
\end{figure}

A few teams used Python simply as a calculator (see Figures~\ref{fig:CT_stud_sample_01} and~\ref{fig:CT_stud_sample_02}), rather than taking advantage of its potential for automation. This tendency to perform calculations without leveraging Python's capabilities for automating repetitive tasks highlights an area for growth in computational thinking, specifically in embracing a more programmatic approach to problem-solving.

Data visualization through graphs (see Figure~\ref{fig:CT_stud_sample_04}) and tables was moderately represented, with several teams recognizing the value of visually presenting their data or findings, although the sophistication of these visualizations varied. Most teams used built-in basic commands, such as~\lstinline{print} (see Figures~\ref{fig:CT_stud_sample_01} and~\ref{fig:CT_stud_sample_02}), for either displaying the output, or perhaps as a debugging tool. The incorporation of physics formulas was nearly universal, yet varied in complexity, with some teams embedding straightforward calculations. Notably, some teams had typed in incorrect formulas, probably due to oversight. 

As mentioned in Section~\ref{sec:V.C(ii)}, due to a possible over-reliance on Python coding, teams paid little attention to significant digits when reporting numerical results—sometimes extending to 17 digits! (see Figure~\ref{fig:CT_stud_sample_01}). Without controlling for precision, students may miss important learning opportunities related to numerical precision, an important aspect in science learning. This example also highlights how science-based thinking and computational thinking are interconnected.

Fewer than half of the teams included adequate comments in their code (see Figure~\ref{fig:python_inductive_coding} in~\hyperref[Appendix C]{Appendix C}), highlighting the need to emphasize this practice in future iterations. Commenting in Python, achieved by placing the \# symbol before a line, provides explanations without affecting the program's execution (see Figure~\ref{fig:CT_stud_sample_03}). This practice is particularly important in group work, as it helps team members follow and understand the code, and ensures that the logic is clear for future reference, especially when the code extends to multiple pages in more advanced tasks.

Overall, students demonstrated a range of CT practices, with an evident use of basic Python libraries and data visualization, but limited attention to numerical precision and debugging. Most teams wrote relatively simple, short programs, suggesting a tendency to stick to basic solutions, perhaps due to time constraints or a lack of programming experience.

\subsubsection{CT -- Deductive Coding}\label{sec:V.F(iii)}

\begin{figure}[htbp]
\caption{A code snippet from Team-13 showing how the team engaged in \textit{brute force} iteration. The team used coding blocks similar to this several times for angles 15$^{\circ}$, 30$^{\circ}$, 60$^{\circ}$, 75$^{\circ}$, and 90$^{\circ}$.}
\fbox{\includegraphics[width=0.96\linewidth]{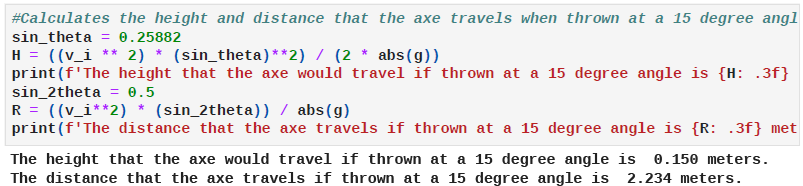}}
\label{fig:CT_stud_sample_02}
\end{figure}

\begin{figure}[htbp]
\caption{A code snippet from Team-01 demonstrating the definition of a custom function and the use of a while loop to update momentum, highlighting the computational thinking aspects of~\textit{Algorithmic Thinking},~\textit{Iteration}, and~\textit{Generalization}. The code is also adequately commented, a good coding practice recommended for beginner coders.}

\fbox{\includegraphics[width=0.96\linewidth]{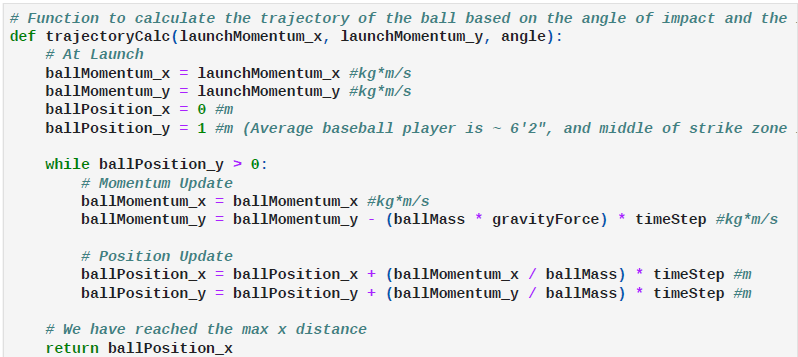}}
\label{fig:CT_stud_sample_03}
\end{figure}

\begin{figure}[htbp]
\caption{A code snippet from Team-04 using~\lstinline{matplotlib} to plot the trajectory of the projectile for different initial angles. This visualization highlights the CT aspects of~\textit{Abstraction} through the representation of physical concepts in a graphical form, and~\textit{Generalization} by adapting the code to display multiple scenarios with varying launch angles.}
\fbox{\includegraphics[width=0.96\linewidth]{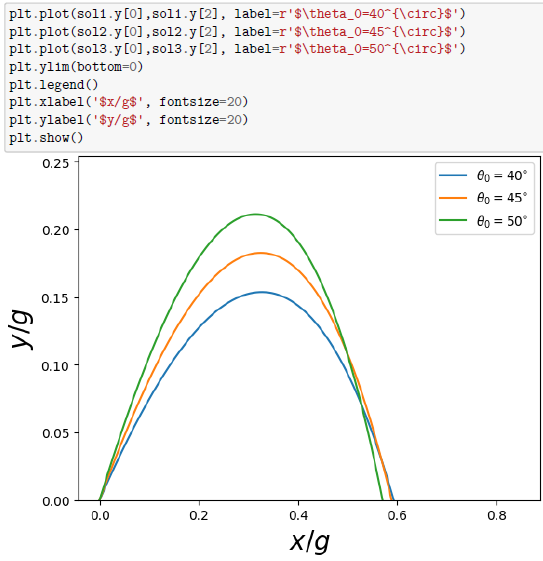}}
\label{fig:CT_stud_sample_04}
\end{figure}

In analyzing students' engagement with computational thinking, we observed how the six elements of Shute {\em et al.}'s framework~\cite{shute2017demystifying} emerged across the fourteen teams’ projects.
\begin{enumerate}

\item\textit{Problem decomposition} manifested in a variety of ways depending on students teams' comfort level with coding in Python. Broadly, teams were able to break down the problem into manageable sub-tasks, often incorporating libraries such as \lstinline{numpy} (\lstinline{np}) for numerical calculations and \lstinline{pandas} (\lstinline{pd}) for data handling. 

\item\textit{Abstraction} was reflected in students' ability to model real-world phenomena by applying appropriate physics principles and concepts, particularly from Newtonian Mechanics. Students also imported modules such as~\lstinline{math} for mathematical operations, and libraries such as~\lstinline{matplotlib} (\lstinline{plt}) (see Figure~\ref{fig:CT_stud_sample_04}) for visualizing data through plots and graphs, further supporting their abstraction processes.

\item\textit{Algorithmic Thinking} emerged from students' engagement with key programming concepts in Python: initializing variables and assigning values to them, applying physics formulas for numerical calculations, and using custom functions (via the~\lstinline{def} statement -- see Figure~\ref{fig:CT_stud_sample_03})
and loop structures (e.g., a~\lstinline{while} loop -- see Figure~\ref{fig:CT_stud_sample_03}) to perform repetitive tasks. 
Functions allowed students to encapsulate specific calculations into reusable code blocks. In addition, students used conditional (\textit{if-then} decision-making) statements throughout their programs.

\item\textit{Debugging} practices were not evident in the final submissions, suggesting that students either did not engage in debugging or likely  did so during the five-week development phase. They may have used the \lstinline{print} statement as a partial debugging tool, though it is less reliable than more advanced methods. While TAs often recommend it for quick insights, its simplicity, universality, and immediate feedback make it a popular choice despite its limitations.

\item\textit{Iteration}, another key element of computational thinking, was explicitly present in a subset of projects through the use of loops (e.g.,~\lstinline{while} -- see Figure~\ref{fig:CT_stud_sample_03}) and other explicit, though evidently less efficient,~\textit{brute force} methods for repeated calculations (see Figure~\ref{fig:CT_stud_sample_02}). This iterative approach played a crucial role in refining students' solutions and automating their calculations.

\item Lastly,~\textit{Generalization} appeared in some teams' use of custom functions (using~\lstinline{def} -- see Figure~\ref{fig:CT_stud_sample_03}) and variables, demonstrating attempts to make their code adaptable to different scenarios. For instance, teams used functions to calculate trajectory parameters, which enabled them to adjust conditions such as angles or initial velocities without altering other parts of their programs.
\end{enumerate}

\subsubsection{CT - Summary and Implications for Instruction}\label{sec:V.F(iv)}

Our dual coding approach, combining inductive and deductive analyses, provided a comprehensive view of students' computational thinking (CT). The inductive analysis highlighted students’ organic coding practices, reflecting their practical, task-driven problem-solving strategies. In contrast, the deductive analysis examined how these practices aligned with established CT principles, as outlined by Shute {\em et al.} While these two approaches appear distinct—one focusing on students' practical, problem-solving practices and the other on established benchmarks—they complemented each other. Together, they provided a deeper understanding of students' CT engagement, revealing both their strengths and challenges in applying CT principles to their work.

The projects demonstrated varying levels of computational thinking, with a small faction of teams showing advanced coding practices in Python, while a majority displayed a more limited approach, often using Python as a basic calculator rather than a programming tool. This range of proficiency suggests a need for scaffolding CT skills, particularly in the areas of debugging and iteration techniques. 

We suggest that future iterations of the course should include more explicit instruction in debugging practices, with increased use of~\lstinline{matplotlib} for creating graphs and tables to help students visualize data and present findings. Additionally, some advanced iteration techniques, such as loop structures, may be be incorporated where appropriate. To further support students, more directed scaffolding for coding activities could be provided, guiding students through the process with incremental steps. This should be done in a way that is mindful of the fact that these are first-year students in an introductory physics course, ensuring that our expectations do not overwhelm them. The focus must remain on physics learning, and any additional coding instruction should complement, rather than detract from, their understanding of core physics concepts. Explicit exercises focused on round-off error, floating-point precision, and significant digits could encourage students to engage critically with the issues of numerical accuracy. A rubric outlining expectations, along with prompt and detailed TA feedback, could also help guide student progress. By gradually introducing these skills in a supportive manner, instructors can better help students develop a more comprehensive and sophisticated approach to computational thinking in physics contexts.

\section{Limitations of the study}\label{sec:VI}

We now outline the limitations identified in our study:

(i) \textbf{Sample size}: The study involved 14 teams comprising 42 students. Although these students were selected from a large-enrollment course of approximately 1,600 students, the sample size remains relatively small.

(ii) \textbf{Local Generalizability within context}: While our goal was to present our findings in sufficient detail to facilitate local transferability (to at least within our own educational settings), we acknowledge the need for larger studies that address the potential for generalizability. While limited in scope, these results help identify emerging trends in how students engage with Engineering Design (ED) problems. Such patterns, though context-bound, offer valuable guidance for refining instructional practices and may inform future studies in similar introductory physics or STEM courses.

(iii) \textbf{Collaborative work}: Since all activities were conducted in teams, the discussions and reports reflect collaborative thinking. As a result, it was not possible to ascertain the individual contributions or thought processes of each student. However, it is important to note that authentic STEM practice is usually a collaborative rather than an individual process.

(iv) \textbf{Selection of data}: Although the project spanned five weeks, only the final week’s submissions were analyzed. Earlier stages of the students' design and problem-solving processes were not included, which might have provided additional insights into the evolution of their thinking.

(v) \textbf{Physical Model constraints}: Due to resource limitations, students developed thought solutions rather than physical models. This lack of tangible prototypes may have influenced the way they approached and expressed their design thinking.

(vi) \textbf{Simplified Computational Thinking}: While computational thinking was included as a component in our framework, we applied a simplified version restricted to python code. A more rigorous implementation, such as that outlined by Shute {\em et al.}, might have yielded different insights.

(vii) \textbf{Impact of ChatGPT}: With ChatGPT introduced in November 2022, it is possible that some students may have used it in their work. Although the lead author, who served as the GTA for the lab section, did not observe any active use of ChatGPT, its usage cannot be ruled out entirely. We were unable to account for the extent to which students might have leveraged the tool, particularly in their Python coding activities. This unexamined factor could have influenced their problem-solving approaches and outcomes.

(viii) \textbf{Non-inclusion of TA feedback}: Feedback provided by the TAs during the lab sessions was not included in our analysis. However, such feedback may have influenced students’ thinking, decision-making, and final outcomes. The absence of this data introduces a limitation, as it leaves certain aspects of the students' learning processes unexplored.

\section{Seeking Qualitative Rigor through Reflexivity}\label{sec:VII}

Given that metacognitive reflection is central to our framework, we saw value in turning that lens on our own work. While some authors~\cite{research_design_cohen_2023, mcpadden_2023} recommend including a~\textit{positionality statement} in qualitative writing, we follow Braund {\em et al.}~\cite{braund_reflexivity} in referring to this section as a~\textit{reflexivity statement}. Their six-way guideline not only shaped the preparation of this statement but also informed our research process more broadly. Since reflexivity is an ongoing requirement throughout the research, an apparent overlap with earlier discussions (see Section~\ref{sec:III.C} and Section~\ref{sec:IV.D}) is inevitable. In many ways, reflexivity extends beyond this section, as our methodological discussions also engage with and reinforce reflective considerations throughout the paper.

In this context, Markham shares a precious insight: 
\begin{quote} 
\textit{Before we can know what we are looking at, we have to know where we’re looking from...}~\cite{markham_reflexivity},~\cite{schuster_reflexivity_2023}
\end{quote}

This perspective aligns with Barth-Cohen {\em et al.} encourage qualitative physics education researchers to include~\textit{positionality statements}, highlighting how their experiences, research roles, and personal knowledge of the study’s topic or domain may have influenced the work~\cite[p.10]{research_design_cohen_2023}. 

Physics, as a predominantly quantitative discipline, demands rigor. Having been deeply rooted in this approach throughout his career as a physics educator, the lead author embraced the challenge of employing qualitative methods in his Physics Education Research (PER). A central question throughout his work has been:~\textit{How do I infuse rigor into my qualitative research?} To this end, he pursued formal training in qualitative methods, obtaining a graduate certificate in qualitative research methodology. The extensive citations in this article reflect this quest for rigor, emphasizing the importance of methodological rigor within qualitative inquiry. Yet, the lead author humbly acknowledges that the pursuit of rigor is an ongoing process.

Beyond formal methodological training, the lead author engaged deeply with the data over an extended period. His prior experience as a physics educator proved invaluable in interpreting student discussions and writing, which, while rooted in the quantitative sciences (physics, mathematics, and python code), required translation into qualitative language for meaningful analysis.

In our approach to qualitative analysis, the lead author—despite being well-versed in technological tools such as NVivo~\cite{nvivo}—has consistently chosen to conduct the majority of the work manually, from transcription and coding to reporting findings. This deliberate choice is rooted in a commitment to staying intimately connected with the data, ensuring that the nuances and complexities of students’ thinking are not abstracted away by automated processes. While undeniably time-intensive, this hands-on approach aligns with Hordge-Freeman’s call to~\textit{``bring your whole self to research''}~\cite{hordge2018bringing}, embracing the deeply immersive and interpretive nature of qualitative inquiry.

In the first paper~\cite{ravi_SWoT_01_2024}, we engaged in a rigorous effort to develop our framework. This process was informed not only by prior research but also by our own direct experiences as educators and educational researchers working with students. Reform documents and research articles, both within and beyond PER, played a critical role in shaping our thinking. Particularly rewarding was the comparative analysis of our framework with those of Dalal {\em et al.} and English. Equally influential were the thoughtful critiques from reviewers of our recently accepted publications. Their insights helped us address key gaps in our work and compelled us to approach this current study with even greater care.

In this second part, our focus shifted to applying the~\textit{WoT4EDP} framework. There were times when we found ourselves wishing for additional data, particularly in the form of physical prototypes, which could have provided deeper insight into students’ thinking. Nevertheless, our findings have been invaluable in informing our pedagogical interventions, highlighting areas for improvement in our own instruction. 

An notable contrast is that the first paper~\cite{ravi_SWoT_01_2024} of this multi-part series is theoretical, while this second part is empirical, both involving the application of qualitative methods. We hope our work has contributed to the ongoing discussion on the use of qualitative methods in Physics Education Research and, more broadly, in Science Education Research~\cite{devetak2010role}.

From data collection to analysis and reporting, we were always reminded by O'Dwyer’s perspective:

\begin{quote}
\textit{There is never going to be one ‘true’ story from a set of qualitative data but the process described above indicates a rigorous attempt at being systematic and reflective.}~\cite[p.406]{odwyer_qlr_messy_intimate} 
\end{quote}

One key question we continuously grappled with throughout the study was our writing style. Gilgun advocates for a more evocative and personal style in qualitative research writing. Perhaps in future work, we will revisit and more fully engage with her call to explore the~\textit{``marvelous range of choices''}~\cite[p.261]{gilgun2005grab} qualitative researchers have.

Among the first articles the lead author read when beginning his journey into qualitative research less than three years ago was~\textit{Getting Started in Qualitative Physics Education Research}~\cite{otero2009getting}. Now, at the conclusion of this paper, he is reminded of the inspiring words of Otero and Harlow:

\begin{quote}
\textit{Qualitative research is never done, but the work you do can be of great value to others. Most of all, have fun!}~\cite[p.61]{otero2009getting}
\end{quote}

\section{Conclusion}\label{sec:VIII}
At the study's outset, we posed three research questions (RQs) to apply our~\textit{WoT4EDP} framework to student data. Here, we summarize our findings for each.
While qualitative findings are not intended for generalization~\cite[p.4]{otero2009getting}, we aimed to explore whether insights from a small group could illuminate broader issues in our context. Our analysis was guided by this objective. In Section~\ref{sec:VIII.A} we review~\hyperref[RQ1]{RQ1}; in Section~\ref{sec:VIII.B} we review~\hyperref[RQ2]{RQ2}; and in Section~\ref{sec:VIII.B} we review~\hyperref[RQ3]{RQ3}. 

\subsection{What do students' ED problem statements reveal?}\label{sec:VIII.A}

In Section~\ref{sec:V.A}, seeking answers for our first research question~\hyperref[RQ1]{RQ1}, we first examined how students framed their problem statements—a crucial step in design-based thinking. 

Key themes include diverse problem selection, creative framing, stakeholder considerations, and real-world relevance. Teams incorporated numerical metrics, constraints, and assumptions, with some showing an entrepreneurial mindset. All problems had connections to physics and mathematics. 

Students would benefit from additional scaffolding to grasp terms like stakeholders, criteria, constraints, and metrics. Since these terms are rarely used in physics classrooms, educators may face challenges in integrating ED-based learning. We hope our work resonates with those seeking to bridge this gap. Our coding sheet in~\hyperref[Appendix A]{Appendix A} can be refined and adapted by educators to develop rubrics for similar activities.

We outlined our challenges in precisely distinguishing between a constraint and a criterion. Despite this, our view is that it may be helpful for students to consider these terms in their discussions. While students demonstrated creativity in framing problems, some teams selected problems that were either overly ambitious or closely aligned with standard textbook examples. 

Framing ill-structured problems remains inherently challenging, particularly in large-enrollment courses often led by TAs (as was the case for the lead author). Professional development workshops on WoT frameworks and ED-based physics problems could provide valuable support in this regard. 

\subsection{How do students engage with design, science, mathematics, and metacognitive reflection in their ED problem?}\label{sec:VIII.B}

Our second research question,~\hyperref[RQ2]{RQ2}, aimed to investigate how students engage with four of the five elements of our framework: design-based thinking (DBT), science-based thinking (SBT), mathematics-based thinking (MBT), and metacognitive reflection (MER).

\subsubsection{Design-Based Thinking (DBT) -- Summary}\label{sec:VIII.B(i)}

Regarding DBT (see Section~\ref{sec:V.B}), key themes include students' awareness of design criteria, constraints, and numerical details, with varying levels of iteration and refinement. Most teams cited references, but safety and economic considerations were less emphasized. While some teams iterated meaningfully, others showed minimal engagement. Trade-offs were not well-articulated, suggesting a need for better guidance. Overall, students demonstrated design thinking aligned with Crilly’s definition~\cite{crilly_design_2024}, balancing creativity and analytical problem-solving.

One area of concern was how students engaged in iterations. In some cases, iterations were overly simplistic, and some teams engaged in only cursory iterations. Although we did not explicitly instruct students to elaborate on the physical dimensions of their design or the mechanics of their contraption, most teams naturally engaged with these aspects, suggesting an intuitive awareness. However, a more intentional focus on these considerations could further enhance their designs. Encouraging students to conduct feasibility studies could also be beneficial, though this would require explicit instruction on what feasibility entails. Having students develop physical models would be ideal, but this is often impractical in large-enrollment courses. An alternative cost effective approach could be to incorporate process-based problems that involve simulations or open-access data analysis.
\subsubsection{Science-Based Thinking (SBT) -- Summary}\label{sec:VIII.B(ii)}

Regarding SBT (see Section~\ref{sec:V.C}), key themes include the use of physics principles and concepts, mention of assumptions and approximations, identification of system and surroundings, reporting results, instances of errors, inaccuracies, and oversights. 

Very few teams applied the angular momentum principle, likely due to the nature of their chosen problem or the fact that this topic is introduced later in the semester—after students had already defined their problem in Week 10. However, given that this is a physics course, not surprisingly, all teams engaged with physics concepts and principles. Here again, our rubric (see~\hyperref[Appendix D]{Appendix D}), though fairly basic, appears to have provided some structure. That said, more guidance is needed on the depth of application of these principles. There were scattered but notable instances of misconceptions or inaccuracies in students' work, which is a crucial issue to address in a physics course. While using the~\textit{Jupyter Notebook} has clear advantages, it also results in students performing calculations without paying attention to numerical precision and units—an oversight that could have serious implications in real-world scenarios. Some teams incorrectly typed their equations, leading to incorrect results, yet they did not seem to recognize the errors. Notably, a few teams wrote their calculations by hand, and while their equations were correctly formulated, they still neglected significant digits. Interestingly, their handwritten work generally avoided excessive numerical precision, unlike those of teams who had~\textit{Jupyter Notebook} outputs. Incorporating elements of the CER (Claim-Evidence-Reasoning) framework~\cite{toulmin2003uses, slavit2021student, mcneill2011claims, siverling2021initiates} could help strengthen the \textit{design-science} connection by encouraging students to articulate and justify their scientific reasoning more explicitly.
\subsubsection{Mathematics-Based Thinking (MBT) -- Summary}\label{sec:VIII.B(iii)}

Regarding MBT (see Section~\ref{sec:V.D}), key themes include how students performed and presented their calculations, and how they reported their findings. 

Most teams were comfortable using algebraic equations, though occasional errors were observed. Only one team employed brute-force calculus, while a few others used it indirectly through the momentum update principle. Students also appeared comfortable generating graphs and tables using Python, with some teams opting to draw graphs by hand. Additionally, we noted instances of qualitative reasoning in their discussions.
Notably, compared to our earlier study~\cite{ravi_prper_2024}, students engaged in more detailed discussions of mathematical ideas. It appears that explicitly prompting students to discuss relevant physics principles may have indirectly encouraged them to integrate mathematical reasoning using physics formulas and principles. 

Overall, in comparison to our earlier study~\cite{ravi_prper_2024}, the rubric (see~\hyperref[Appendix D]{Appendix D}) we provided seems to have been beneficial. This highlights the need for further research on designing more effective rubrics that can better support and enhance student engagement with mathematical reasoning in physics-based design problems.
\subsubsection{Metacognitive Reflection (MER) -- Summary}\label{sec:VIII.B(iv)}
Regarding MER (see Section~\ref{sec:V.E}), we took a slightly different approach in selecting the portion of data, focusing solely on the final question in the report, where we had prompted students to engage in what we \textit{retrospectively} viewed as a reflective evaluation. However, our prompt appears to have been ineffective in guiding students. Most students seemed unclear about what was expected, and while they did produce responses, the task did not meaningfully engage their time. This is not a judgment on the students but rather a reflection on the need for better scaffolding on our part. This was a significant lesson for us—while we were eager to give students agency, we lacked sufficient planning in structuring the reflection task. We will address this issue in the upcoming semester by refining our prompts and instructional approach. As much as we may critique our own implementation, this serves as a reminder that designing effective instructional activities is challenging, underscoring the importance of continued research in education.

\subsection{In what ways do students demonstrate computational thinking in their ED challenges?}\label{sec:VIII.C}
As for our third research question~\hyperref[RQ3]{RQ3}, we chose to investigate computational thinking (restricted to students' Python codes), separately because it is a novel element in our framework that has not been explored in other frameworks. In fact, we did not find empirical studies that discuss computational thinking (CT) specific to use of python coding. The importance we attached to this element is evident in our adoption of both inductive and deductive approaches—an intentionally elaborate exercise (see Section~\ref{sec:V.F}).

Our ED-based project provided an ideal context for students to demonstrate their computational skills, particularly within the~\textit{Jupyter Notebook} environment, which facilitates such engagement. We were impressed by how some teams used Python coding to extend their problem-solving efforts. This is despite that we did not set high expectations for students in this task—even a simple tweaking of a few coding cells from prior activities was deemed sufficient. While a few teams exceeded our expectations, there were some teams that did not engage deeply with Python, and some teams chose to perform all computations manually.

Investigating this research question was an enlightening experience, as it revealed the diverse ways in which students approach computational thinking. These insights will serve as a valuable guide for refining our approach in the next iteration of the course. Additionally, we believe our detailed coding table may provide educators with useful insights for structuring instructional activities that incorporate Python coding.

Notably, in a recent study, Gambrell and Brewe highlight that in the context of CT in physics,~\textit{``there is still a lot of research to be done in many of the specific physics courses''}~\cite[p.5]{CT_brewe_gambrell}. We hope we have meaningfully contributed to the discussion on computational thinking.

\section{Implications}\label{sec:IX}
We believe this work holds significance for researchers, educators, and even students. For researchers in Physics Education Research (PER), this study provides a detailed application of a SWoT framework in physics classrooms, demonstrating its utility in Engineering Design (ED)-based instruction through qualitative analysis. In the broader landscape of science education research, we aim to contribute to ongoing discussions on Ways of Thinking (WoT) frameworks. Beyond science and education, we also believe our work may resonate with qualitative methods researchers. 

Having been a teacher before transitioning to graduate school, the lead author is acutely aware of the challenges in designing instructional materials, particularly within STEM contexts. Time constraints are a persistent concern for educators, as any teacher would attest. Teachers need frameworks that are both simple and adaptable, along with guidance on their implementation. They also need to feel empowered to make instructional decisions at a local level to further their immediate pedagogical goals. This recognition has informed our deliberate emphasis on flexibility—both in the development of WoT frameworks and in their application. Perhaps this work might even inspire a high school teacher to transition to graduate school and contribute to the field.

A practical benefit of exploring various WoT frameworks is that teachers can enhance their rubrics, incorporating diverse perspectives to better evaluate and support student performance. Awareness of WoT frameworks can also help students reflect on and refine their own thinking, making their learning process more intentional. Rubrics informed by WoT frameworks can clarify evaluation criteria, guiding students to focus on crucial aspects of their work.

While our primary audience includes researchers and educators, we believe this work also offers students valuable examples and insights into how their peers think. Studies highlight the potential of~\textit{peer learning}~\cite{peer_topping_01122005} to~\textit{``foster fertile instructional dialogues among peers''}~\cite[p.4]{peer_bozzi_2021}, and we hope student readers will find meaning in the thought processes of their unseen peers.

Ultimately, working on the second paper of this multi-part series has been an immense learning experience, and we remain humbled by the complexity of studying human thinking. At the end of the day, our \textit{WoT4EDP} framework is \textit{just a model}. Likewise, our demonstration of qualitative techniques for analyzing student data is only one possible approach. As the saying goes:

\begin{quote} 
\textit{All models are wrong, but some are useful.}~\cite[p.202]{box1979science}
\end{quote}

\section{Future Work}\label{sec:X}

In the course of this study, we occasionally referred to interconnected ways of thinking, but we did not delve into the details. In the next part of this series, we will explore how the interconnections between various \textit{Ways of Thinking} unfold. After having examined these elements separately, discussing~\textit{Integrative Thinking} (see Figure~\ref{fig:wot_framework}) may seem contradictory. It is our view that by intentionally focusing on each element during instruction, we can foster deeper integration, ultimately leading to more effective solutions. Naturally, this discussion will also address how~\textit{Integrative Thinking} itself may be defined.

Another potential avenue for exploration is how students engage in~\textit{argumentation}. Following Slavit {\em et al.}~\cite{slavit2022analytic}, we are considering applying Toulmin’s \textit{Claim–Evidence–Reasoning} (C–E–R) framework~\cite{toulmin2003uses, mcneill2011claims} to analyze how students construct and justify their reasoning.

\section{Acknowledgments}
We gratefully acknowledge the insightful and encouraging comments of anonymous referees on our earlier publications that contributed to what we believe is a more coherent, detailed, and well-structured paper. We owe our gratitude to all the authors referenced, for they provided us with innumerable moments of much needed inspiration. ChatGPT-4o and Perplexity were employed by the lead author to~\textit{wordsmith} passages, and as a~\textit{learning partner}. A special thanks to Amir Bralin for designing the learning materials, with particular emphasis on scaffolding the Python-based activities, preparing the~\textit{Jupyter Notebooks} for the laboratory reports, and carefully reviewing our analysis of computational thinking in Section~\ref{sec:V.F}. The lead author assumes full responsibility for any errors and omissions.

This work is supported in part by U.S. National Science Foundation grant DUE-2021389. The opinions expressed are those of the authors and not of the Foundation. 
\clearpage
\bibliography{references}
\clearpage
\appendix
\onecolumngrid

\appendix
\onecolumngrid

\section*{Appendix A}
\label{Appendix A}

\FloatBarrier 
In this appendix, we provide the coding sheet with the coded student data for the problem statements of all 14 teams.  See Section~\ref{sec:V.A} for further details.
\begin{figure*}[htbp]
    \captionsetup{position=above} 
    \caption{Coding sheet for the ED problem statement framed by student teams. `1' denotes existence of the code. Code details are provided at the bottom of the table.}
    \label{fig:prob_stmt_code_01}
    \centering
    \fbox{\includegraphics[width=0.9\textwidth]{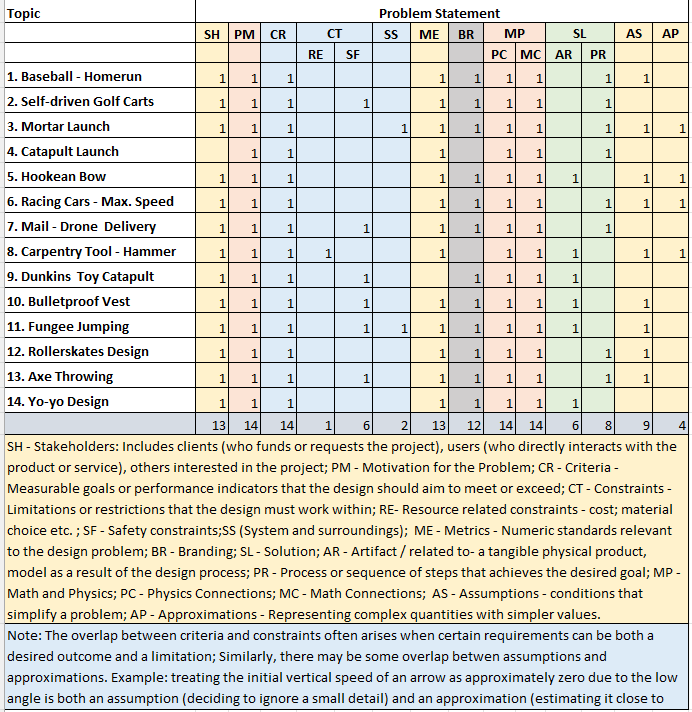}} 
\end{figure*}

\section*{Appendix B}
\label{Appendix B}

\FloatBarrier 
In this appendix, we provide the coding sheet with the coded student data on design-based thinking (DBT) and science-based thinking (SBT) for all 14 teams. See Sections~\ref{sec:V.B} and~\ref{sec:V.C} for further details.
\begin{figure*}[htbp]
    \captionsetup{position=above} 
    \caption{Code Summary for Design-Based Thinking (DBT). See Section~\ref{sec:V.B}}
    \label{fig:DBT_code_summary}
    \centering
    \fbox{\includegraphics[width=0.8\textwidth]{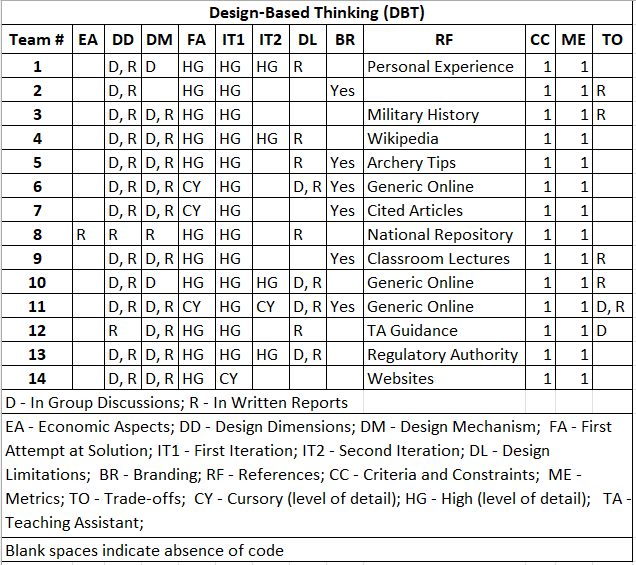}} 
\end{figure*}

\begin{figure*}[htbp]
    \captionsetup{position=above} 
    \caption{Code Summary for Science-Based Thinking (SBT). See Section~\ref{sec:V.C}}
    \label{fig:SBT_code_summary}
    \centering
    \fbox{\includegraphics[width=\textwidth]{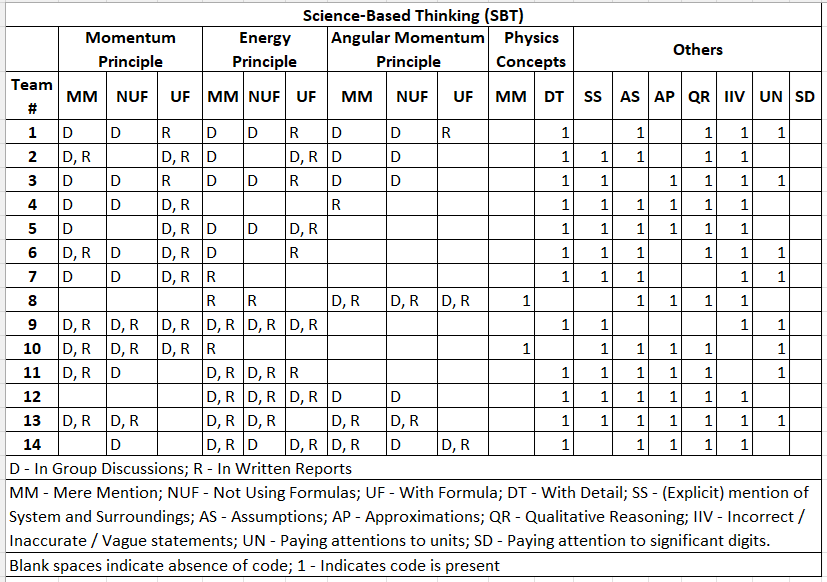}} 
\end{figure*}

\section*{Appendix C}
\label{Appendix C}

\FloatBarrier 
In this appendix, we provide the coding sheet with the coded student data on teams' Python coding. Our analysis employs both inductive and deductive coding approaches within the computational thinking (CT) framework. See Section~\ref{sec:V.F} for further details.

\begin{figure*}[htbp]
    \captionsetup{position=above} 
    \caption{Inductive coding summary for the Python codes developed by student groups. A `1' indicates the presence of the code. See Section~\ref{sec:V.F(ii)} for details.}
    \label{fig:python_inductive_coding}
    \centering
    \fbox{\includegraphics[width=\textwidth]{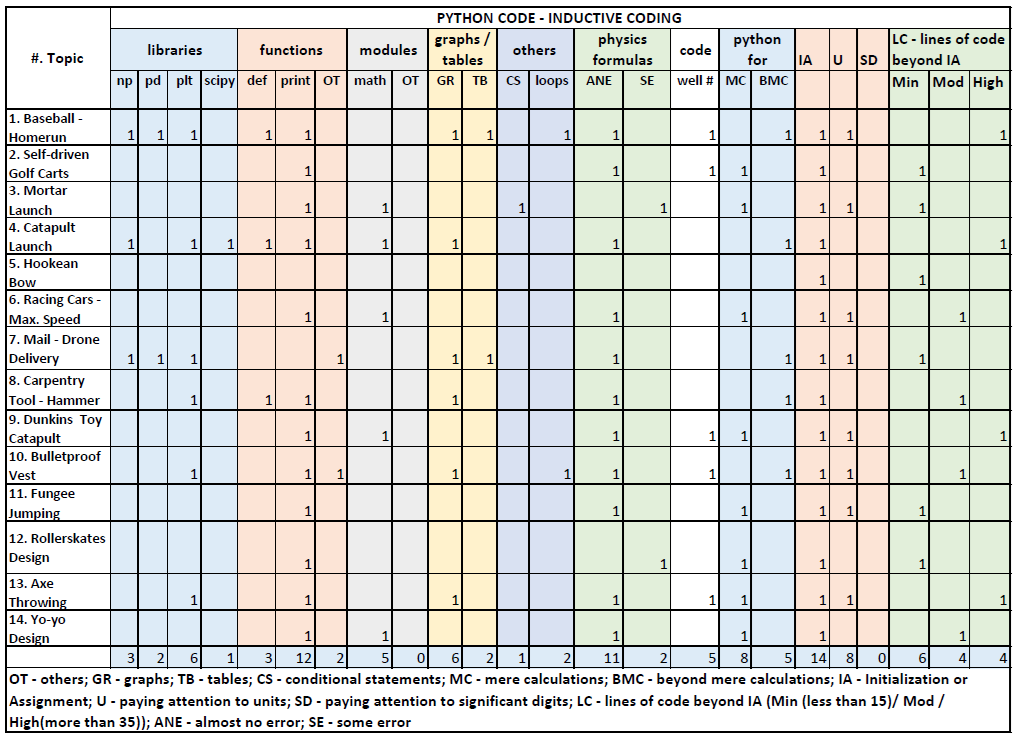}} 
\end{figure*}

\FloatBarrier 

\begin{table}
\centering
\caption{Deductive coding summary based on the six CT facets of Shute {\em et al.} Each of the 14 teams was coded separately according to this framework. This table provides a broad representative summary of all the teams taken as a whole. See Section~\ref{sec:V.F(iii)} for details.}
\small
\begin{tabular}{p{1.5cm}p{2.3cm}p{2.3cm}p{2.3cm}p{2.3cm}p{2.3cm}p{2.3cm}}
\toprule
\textbf{CT facets} & \textbf{Problem \newline{Decomposition}} & \textbf{Abstraction} & \textbf{Algorithms} & \textbf{Debugging} & \textbf{Iteration} & \textbf{Generalization} \\
\midrule
\textbf{Definition} & Breaking down complex problems or systems into manageable, functional parts that collectively form the whole. & Extracting system essence: data analysis, pattern recognition, and modeling. & Designing logical instructions: algorithm design, parallelism, efficiency, and automation. & Identifying, locating, and fixing errors when solutions don't work as intended. & Repeating and refining design processes to achieve optimal results. & Applying computational thinking skills across various situations and domains for effective problem-solving. \\
\midrule
\textbf{Summary of findings across all teams} & Problem decomposition is evident across most projects, with students breaking down complex challenges into more manageable components. & Abstraction is reflected in how teams model real-world phenomena by applying simplified physics principles, concepts, and ideas. & Algorithm design, whether implicit or explicit, is present in nearly every project. & Debugging practices not evident in final submissions, likely occurred during development. Print statements, while limited, are popular for quick insights due to simplicity and effectiveness. & Iteration is demonstrated explicitly in some projects through the use of loops, specific functions, or repeated 'brute force' calculations. & Generalization is often achieved by employing functions and variables, making the code adaptable to different scenarios. \\
\bottomrule
\label{tab:deductive_CT_coding}
\end{tabular}
\end{table}

\section*{Appendix D}
\label{Appendix D}
This appendix presents the rubric given to students for the student-generated ED task.
\FloatBarrier 
\begin{figure*}[htbp]
    \captionsetup{position=above} 
  \caption{Page 1 of the rubric for the student-generated ED problem. Since students worked within a \textit{Jupyter Notebook} environment for computational analysis, the task was titled a Computational Physics Essay (CPE).}
    \label{fig:cpe_rubric}
    \centering
    \fbox{\includegraphics[width=0.8\textwidth]{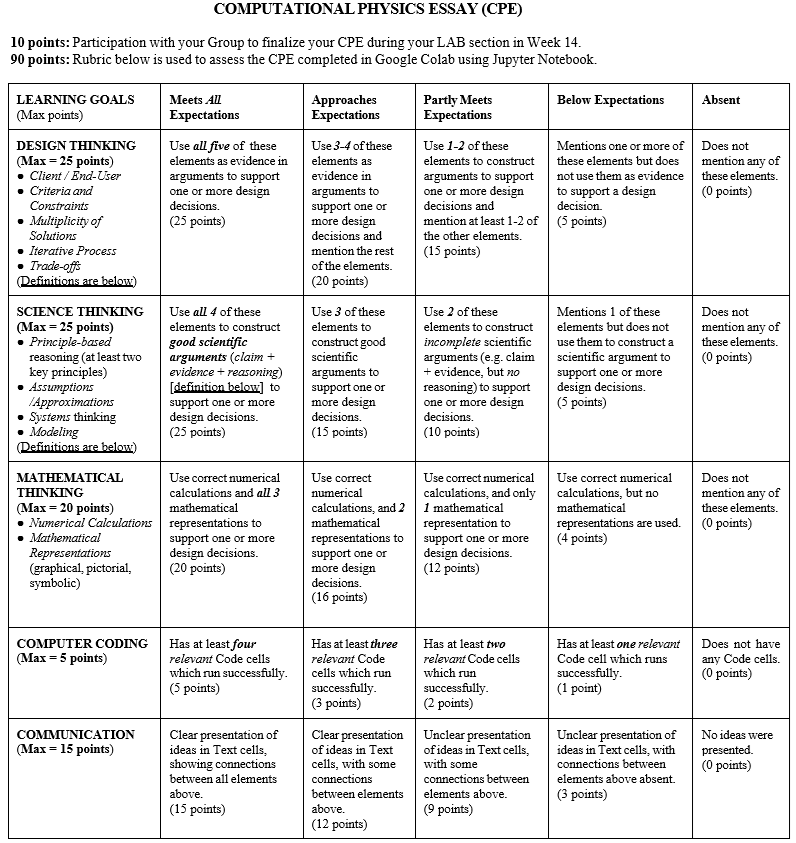}} 
\end{figure*}

\begin{figure*}[htbp]
    \captionsetup{position=above} 
    \caption{Page 2 of the rubric for the student-generated ED problem. }
    \label{fig:cpe_rubric_page_2}
    \centering
    \fbox{\includegraphics[width=0.8\textwidth]{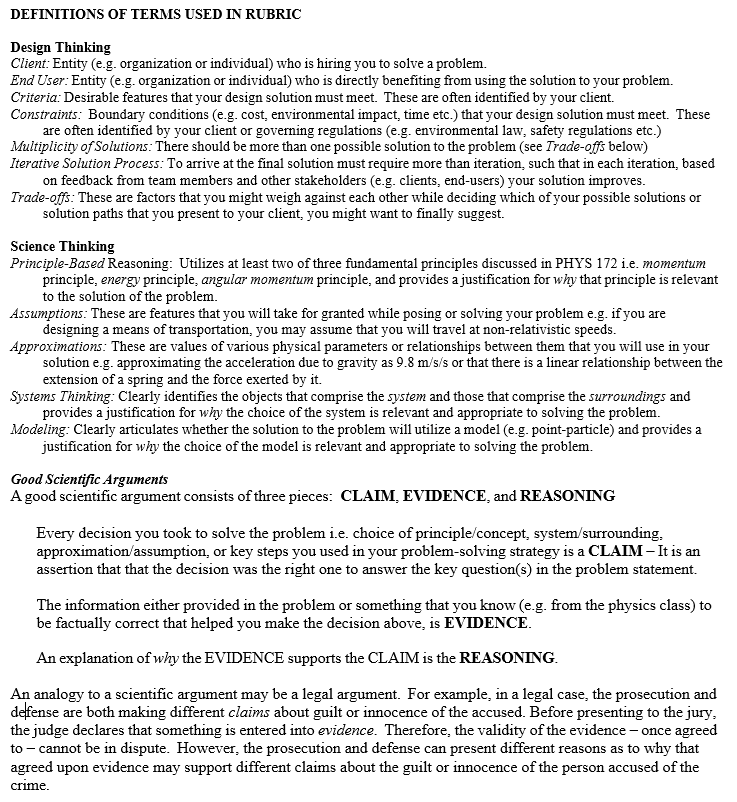}} 
\end{figure*}

\end{document}